\documentclass{aa}  
\usepackage{graphicx}
\usepackage[T1]{fontenc}
\usepackage{txfonts}
\usepackage{epstopdf}
\newcommand{\asec}{$^{\prime\prime}$}

\def\SigmaH2{$\Sigma $(${\rm H_2}$)}
\def\r1415{$^{14}$N/$^{15}$N}

\def\H{N$_{2}$H$^{+}$}

\def\15N{$^{15}$NNH$^+$}
\def\N15{N$^{15}$NH$^+$}

\def\HCOp{\mbox{HCO$^+$}}
\def\HCOpI{\mbox{H$^{13}$CO$^+$}}

\def\HCNHp{\mbox{HCNH$^+$}}
\def\H13CN{\mbox{H$^{13}$CN}}
\def\HII{H{\sc ii}}

\def\kms{\mbox{km~s$^{-1}$}}
\def\cmc{cm$^{-3}$}
\def\cmq{cm$^{-2}$}

\def\solm{\mbox{M$_\odot$}}

\def\Ntot{$N_{\rm tot}$}
\def\Tex{\mbox{$T_{\rm ex}$}}

\def\Tk{\mbox{$T_{\rm k}$}}

\def\kms{km\,s$^{-1}$}

\def\Td{$T_{\rm dust}$}

\begin{document} 

   \title{First survey of HCNH$^+$ in high-mass star-forming cloud cores}
   
   \author{F. Fontani
          \inst{1,2}
          \and
          L. Colzi\inst{3,1}
          \and
          E. Redaelli\inst{2}
          \and
          O. Sipil\"{a}\inst{2}
          \and
          P. Caselli\inst{2}
          }

   \institute{INAF-Osservatorio Astrofisico di Arcetri, Largo E. Fermi 5, I-50125, Florence, Italy\\
            \email{francesco.fontani@inaf.it}
            \and
            Centre for Astrochemical Studies, Max-Planck-Institute for Extraterrestrial Physics, Giessenbachstrasse 1, 85748 Garching, Germany
            \and
            Centro de Astrobiolog\'ia (CSIC-INTA), Ctra. de Ajalvir Km. 4, Torrej\'on de Ardoz, 28850 Madrid, Spain
             }

 \date{Received 25 february 2021; accepted 14 may 2021}

 
 \abstract
   {Most stars in the Galaxy, including the Sun, were born in high-mass
   star-forming regions. It is hence important to study the chemical processes in these regions
   to better understand the chemical heritage of both the Solar System and most stellar systems 
   in the Galaxy.}
   {The molecular ion \HCNHp\ is thought to be a crucial species in ion-neutral astrochemical reactions, but so far
   it has been detected only in a handful of star-forming regions, and hence its chemistry is poorly known.}
   {We have observed with the IRAM-30m Telescope 26 high-mass star-forming cores
   in different evolutionary stages in the $J=3-2$ rotational transition of \HCNHp.}
   {We report the detection of \HCNHp\ in 16 out of 26 targets. This represents the largest
   sample of sources detected in this molecular ion so far. The fractional abundances of \HCNHp\
   with respect to H$_2$, [\HCNHp], are in the range $0.9 - 14 \times 10^{-11}$, and the highest values 
   are found towards cold starless cores, for which [\HCNHp] is of the order of $10^{-10}$. 
   The abundance ratios [\HCNHp]/[HCN] and [\HCNHp]/[\HCOp] are both $\leq 0.01$ for all objects 
   except for four starless cores, for which they are well above this threshold. These sources have the lowest 
   gas temperature and average H$_2$ volume density in the sample.
   Based on this observational difference, we run two chemical models, a "cold" one and a "warm" one, which 
   attempt to match as much as possible the average physical properties of the cold(er) starless cores and
   of the warm(er) targets. The reactions occurring in the latter case are investigated in this work for the first time.
   Our predictions indicate that in the warm model \HCNHp\ is mainly produced by reactions with HCN and 
   HCO$^+$, while in the cold one the main progenitor species of \HCNHp\ are 
   HCN$^+$ and HNC$^+$.}
   {The observational results indicate, and the model predictions confirm, that the chemistry of \HCNHp\
   is different in cold/early and warm/evolved cores, and the abundance ratios [\HCNHp]/[HCN] and 
   [\HCNHp]/[\HCOp] can be a useful astrochemical tool to discriminate between different evolutionary 
   phases in the process of star formation.}


 \maketitle
%

\section{Introduction}
\label{intro}

\begin{table*}
\begin{center}
\caption{Line parameters obtained from Gaussian fits (Cols. 2--6), and total column densities (beam averaged)
of \HCNHp\ obtained from the spectra of the $J=3-2$ line (Fig.~\ref{fig:spectra}) as explained in Sect.~\ref{res}.
The assumed excitation temperatures, in Col. 7, are the kinetic temperatures listed in Fontani et al.~(\citeyear{fontani11}),
and are the same used in Colzi et al.~(\citeyear{colzi18a}) to derive the column densities of the HCN isotopologues.
The last column shows the \HCNHp\ abundances relative to H$_2$, [\HCNHp], computed from the H$_2$ column
densities given in \citet{fontani18}. Because these were averaged values within an area of angular dimension 28\asec,
the given [\HCNHp] are also average values within 28\asec.}
\begin{tabular}{ccccccccc}
\hline \hline
 source   &  $\int T_{\rm MB}{\rm d}V$ &   $V_{\rm p}$    &   FWHM    &  $T_{\rm MB}^{\rm p}$ & 1$\sigma$ & $T_{\rm ex}$ & \Ntot (\HCNHp) & [\HCNHp] \\
               &  K \kms\                             &    \kms\                    &   \kms\       &           K                             &     K               &     K        & $\times 10^{13}$ \cmq\ & $\times 10^{-11}$ \\
\hline
\multicolumn{9}{c}{HMSCs} \\
\hline
I00117--MM2      &    $\leq 0.08$     &      --     &        2.5$^{(a)}$    &     $\leq 0.03$    &    0.01   &  14   &  $\leq 0.94$  &  --$^{(c)}$ \\
AFGL5142--EC$^{(w)}$  &    0.10(0.03)   &    --2.8(0.3)      &      2.3(0.8)     &     0.041   &   0.014    &  25   &  1.0(0.5) & 1.4(1.0) \\
05358--mm3$^{(w)}$      &    0.11(0.02)   &    --16.1(0.3)     &      3.0(0.7)     &     0.035  &    0.01    &  30   &  1.2(0.5)  & 1.7(1.0) \\
G034--G2       &   0.21(0.03)    &    42.9(0.3)       &     4.3(0.6)      &       0.05    &    0.013    &   16  &   2.3(0.8) & 8.6(4.9) \\
G034--F2       &   0.12(0.02)    &    58.0(0.2)       &     2.5(0.7)      &     0.045    &      0.013    &   16  &   1.3(0.5) & 12(8) \\
G034--F1       &    0.10(0.02)   &     57.0(0.3)      &      2.5(0.7)     &      0.037   &      0.012    &  16   &   1.1(0.5) & 14(9)  \\
G028--C1     &    0.27(0.02)   &     79.5(0.7)      &      1.7(0.2)     &      0.150     &    0.015    &  17   &    2.8(0.8) & 12(6) \\
I20293--WC$^{(b)}$    &    0.05(0.01)   &     7.2(0.2)      &      1.4(0.4)     &      0.03   &     0.011    &  17   &     0.5(0.2) & 1.0(0.7) \\
I22134--G$^{(w)}$      &    $\leq 0.06$    &    --      &     2.5$^{(a)}$     &      $\leq 0.024$   &   0.008    &  25   &  $\leq 0.61$ & $\leq 2.6$ \\
I22134--B      &    $\leq 0.07$   &    --  &         2.5$^{(a)}$     &       $\leq 0.03$   &     0.01    &  17   &   $\leq 0.74$  &  $\leq 5.7$ \\
\hline
\multicolumn{8}{c}{HMPOs} \\
\hline
00117--MM1     &    $\leq 0.07$     &      --              &       1.6$^{(a)}$     &      $\leq 0.039$   &   0.013    &  20   &  $\leq 0.67$ & --$^{(c)}$ \\
AFGL5142--MM  &    0.09(0.02)   &    --2.3(0.2)      &      1.9(0.5)     &      0.045   &   0.013    &  34   &  1.0(0.4)  & 1.5(1.0) \\
05358--mm1      &    0.09(0.02)   &    --16.1(0.2)      &      1.4(0.4)     &      0.06   &    0.018    &  39   &  1.0(0.5)  & 1.9(1.3) \\
18089--1732     &     0.18(0.07)    &     32.5(0.3)       &      2.3(0.6)     &      0.07   &   0.020    &  38   &  2(1)      & 3.2(2.6) \\
18517+0437     &   $\leq 0.06$     &      --    &        1.6$^{(a)}$   &      $\leq 0.033$  &    0.011   &  40   &  $\leq 0.69$ & $\leq 1.3$ \\
G75--core         &    $\leq 0.07$     &      --    &        1.6$^{(a)}$    &      $\leq 0.042$  &    0.014  &  96   &  $\leq 1.48$ & $\leq 5.2$ \\
I20293--MM1    &    0.09(0.02)   &     5.6(0.1)      &      1.5(0.2)     &      0.058   &    0.013   &   36  &   1.0(0.4)   & 3.2(2.0) \\
I21307            &  $\leq 0.03$      &       --          &      1.6(0.3)     &      $\leq 0.025$   &    0.009   &   21  &   $\leq 0.30$ & $\leq 1.5$ \\
I23385            &     $\leq 0.07$    &      --            &        1.6$^{(a)}$        &       $\leq 0.039$   &   0.013   &  37   &  $\leq 0.77$  & $\leq 4.9$ \\
\hline
\multicolumn{8}{c}{UCHIIs} \\
\hline
G5.89--0.39      &   0.97(0.04)    &    8.1(0.6)   &      2.5(0.1)     &       0.368      &    0.024   &  32   &  10(3)  & 2.9(1.5) \\
I19035--VLA1   &   $\leq 0.11$     &     --          &      2.4$^{(a)}$          &     $\leq 0.042$    &   0.014    &  39   &  1.3(0.3) & 5.4(2.4)  \\
19410+2336     &   0.08(0.01)    &    22.5(0.1)   &      1.4(0.3)     &      0.054        &   0.010    &  21   &  0.8(0.3) & 0.9(0.5) \\
ON1                 &   0.22(0.2)    &    11.2(0.1)       &     3.2(0.3)      &     0.065       &   0.010    &   26  &   2.2(0.6) & --$^{(c)}$ \\
I22134--VLA1  &    0.08(0.01)   &    --18.0(0.1)      &      1.8(0.3)     &      0.042   &   0.009    &  47   &  1.0(0.4) & 7.3(4.4) \\
23033+5951$^{(b)}$     &    0.10(0.02)   &    --53.4(0.4)      &      3.1(1)     &    0.031   &    0.011    &  25    &  1.0(0.5) & 2.0(1.4) \\
NGC7538--IRS9  &    $\leq 0.08$     &     --     &       2.4$^{(a)}$            &     $\leq 0.033$    &   0.011    &  32   &  $\leq 0.09$  & $\leq 1.1$ \\
\hline
\end{tabular}
\end{center}
$^{(a)}$ fixed FWHM assumed to compute the upper limit on \Ntot (\HCNHp), obtained as the average value of the detected lines in the corresponding evolutionary group;
$^{(b)}$ tentative detections;
$^{(c)}$ column density of H$_2$ not available;
$^{(w)}$ warm HMSCs. The other HMSCs are classified as cold (or quiescent. See Fontani et al.~\citeyear{fontani11}).
\label{tab:results}
\end{table*}

It is now clear that the Sun and most stars in the Milky Way are born in rich clusters 
including, or close to, high-mass stars (e.g.~Carpenter~\citeyear{carpenter00}, Pudritz~\citeyear{pudritz02},
Adams~\citeyear{adams10}, Rivilla et al.~\citeyear{rivilla14}, Lichtenberg et al.~\citeyear{lichtenberg19}).
Therefore, the study of the chemical content and evolution of high-mass star-forming regions can 
give us important information about the chemical heritage of both the Solar System and most stars 
in the Milky Way. Despite the importance of studying the chemistry of high-mass star-forming cloud cores 
(i.e.~compact structures with mass $\sim 10 - 100$ \solm, which have the potential to form single high-mass 
stars and/or clusters), an evolutionary classification is not yet clear due to both observational and theoretical 
problems (e.g.~Beuther et al.~\citeyear{beuther07}, Tan et al.~\citeyear{tan14}, Motte et al.~\citeyear{motte18}, 
Padoan et al.~\citeyear{padoan20}).

Several attempts to empirically give an evolutionary classification of high-mass star-forming cores have been 
proposed, which can all be tentatively summarised in three coarse phases: (1) high-mass starless cores 
(HMSCs), i.e. dense infrared-dark cores characterised by low temperatures ($\sim 10-20$~K) and high 
densities ($n\geq 10^{4}-10^{5}$~\cmc), and without clear signs of on-going star formation like strong
protostellar outflows and masers;
(2) high-mass protostellar objects (HMPOs), i.e. collapsing cores with evidence of one (or more) deeply 
embedded protostar(s), characterised typically by higher densities and temperatures ($n\simeq 10^6$~\cmc, $T\geq 20$~K);
(3) ultra-compact \HII\ regions (UCHIIs), i.e. Zero-Age-Main-Sequence stars associated with an expanding 
\HII\ region, whose surrounding molecular cocoon ($n\geq 10^5$~\cmc, $T\sim 20 - 100$~K) is affected physically
and chemically by its progressive expansion. 

Regardless of the evolutionary stage, high-mass star-forming cores are characterised by a complex and rich 
chemistry (e.g.~Fontani et al.~\citeyear{fontani07}, Bisschop et al.~\citeyear{bisschop07}, Foster et al.~\citeyear{foster11}, 
Belloche et al.~\citeyear{belloche13}, Vasyunina et al.~\citeyear{vasyunina14}, Coletta et al.~\citeyear{coletta20}).
The new generation telescopes have provided a growing amount of observational results 
including, thanks to their high sensitivity, the detection of "rare" species (i.e. species with fractional abundance 
with respect to H$_2$ smaller than $\sim 10^{-10}$). These species can have important 
implications not only for our understanding of the still mysterious process of high-mass star formation, but also 
for the chemistry that the primordial Solar System might have inherited from its birth environment
(e.g.~Beltr\'an et al.~\citeyear{beltran09}, Fontani et al.~\citeyear{fontani17}, Ligterink et al.~\citeyear{ligterink20}, 
Rivilla et al.~\citeyear{rivilla20}, Mininni et al.~\citeyear{mininni20}). 
Among these rare species, protonated hydrogen cyanide, \HCNHp\ (or iminomethylium), is important in astrochemistry
because it is thought to be the main precursor of HCN and HNC. They are both among the most abundant 
species in star-forming regions, and believed to have a high pre-biotic potential (Todd \& \"{O}berg~\citeyear{teo20}). 
Despite its importance, and after the first discovery towards SgrB2 (Ziurys \& Turner~\citeyear{zet86}), 
\HCNHp\ has been detected so far in a handful of other star-forming regions: the TMC-1 dark cloud, the DR21(OH) 
\HII\ region (Schilke et al.~\citeyear{schilke91}), and the low-mass pre--stellar core L1544 
(Qu\'enard et al.~\citeyear{quenard17}).

In this work, we report 16 new detections of \HCNHp\ towards 26 high-mass star-forming cores almost
equally divided into HMSCs, HMPOS, and UCHIIs. Sect.~\ref{obs} describes the sample and the
observational dataset;
Sect.~\ref{res} presents the observational results, which we discuss in Sect.~\ref{discu}; in Sect.~\ref{model} 
we describe a chemical model with which we interpret the observational results, and in Sect.~\ref{conc}
we summarise our main findings.

\begin{figure}
\includegraphics[width=8.5cm]{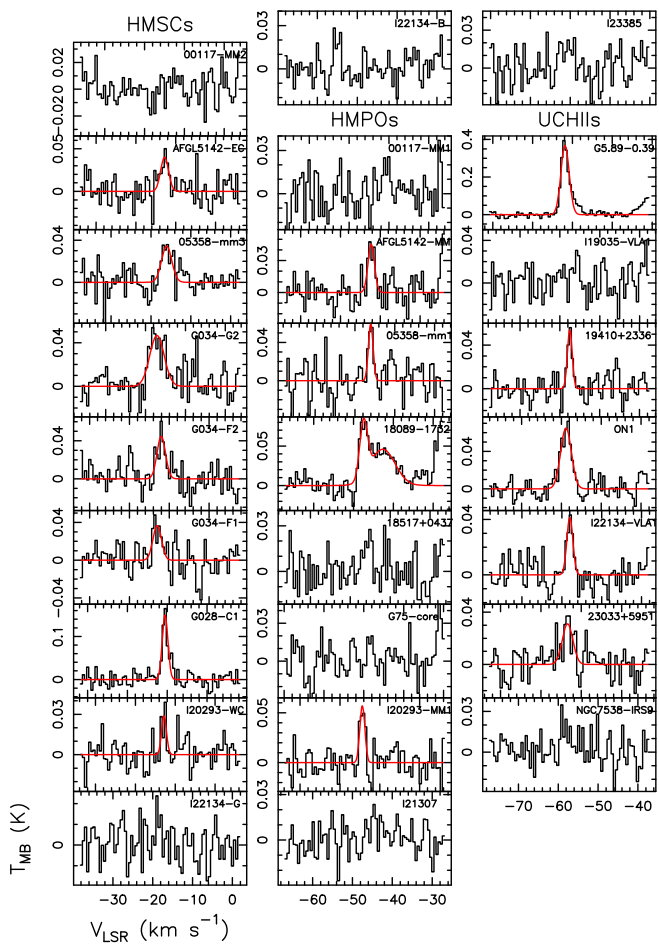}
      \caption{Observed spectra of \HCNHp $J=3-2$ obtained with the IRAM-30m telescope.
      For each spectrum, the velocity interval shown on the x-axis is $\pm 20$~\kms\ around the systemic 
      velocity listed in Table A.1 of \citet{fontani11}.
      The red curves represent Gaussian fits to the detected lines. For 18089--1752, a two-Gaussian fit
      was performed.
              }
         \label{fig:spectra}
\end{figure}

\begin{figure*}
{\includegraphics[width=6cm]{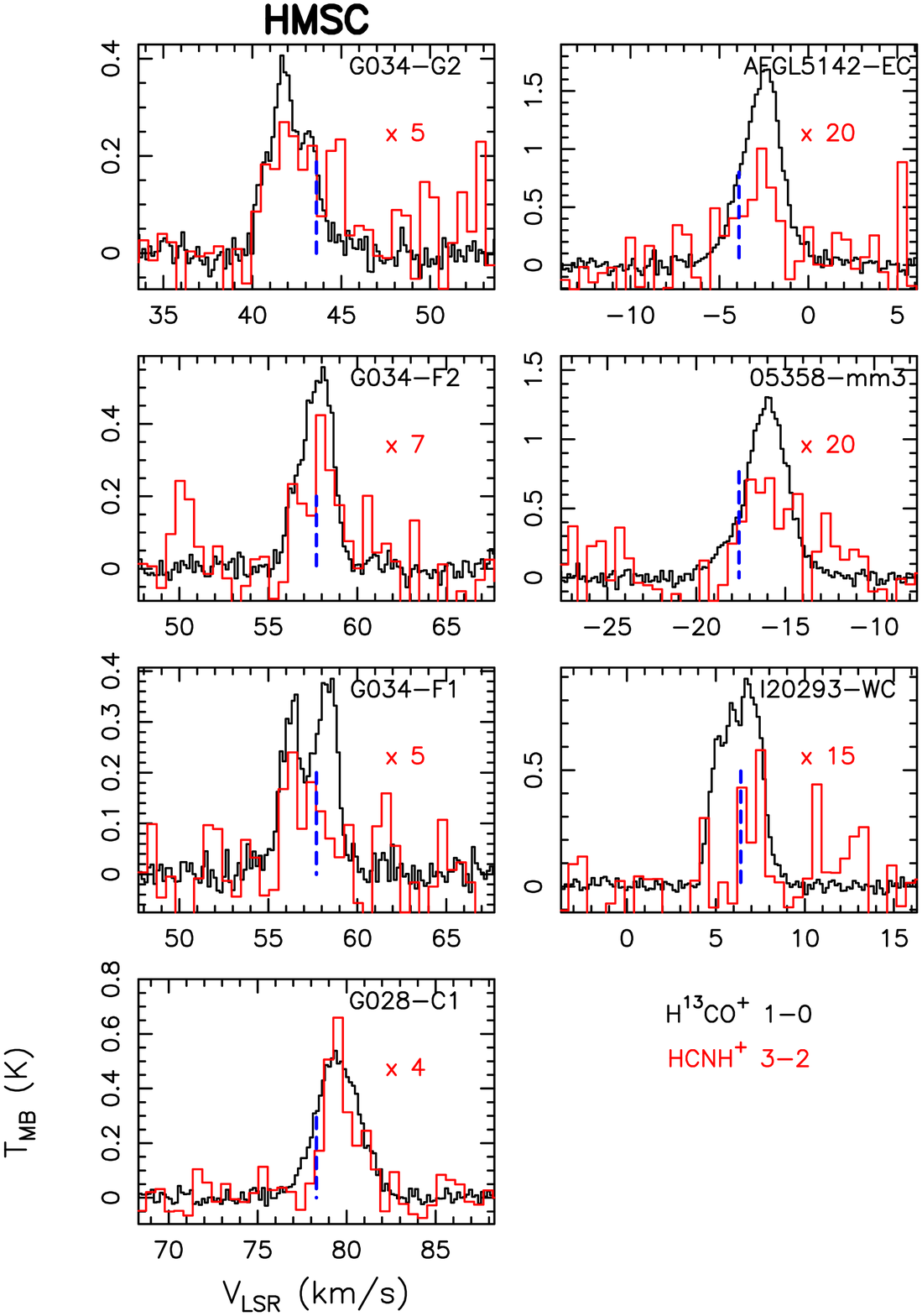}
\includegraphics[width=6cm]{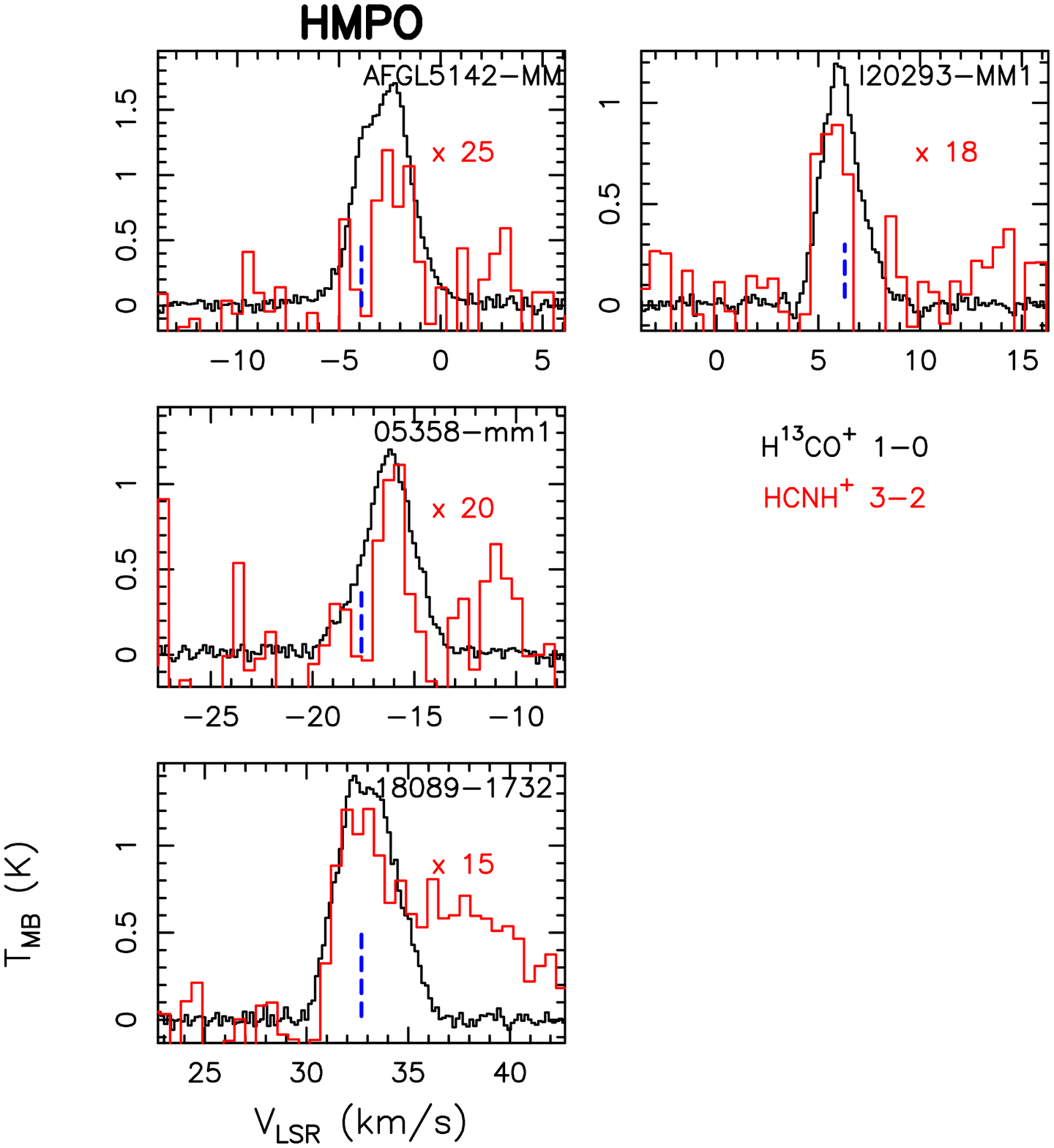}
\includegraphics[width=6cm]{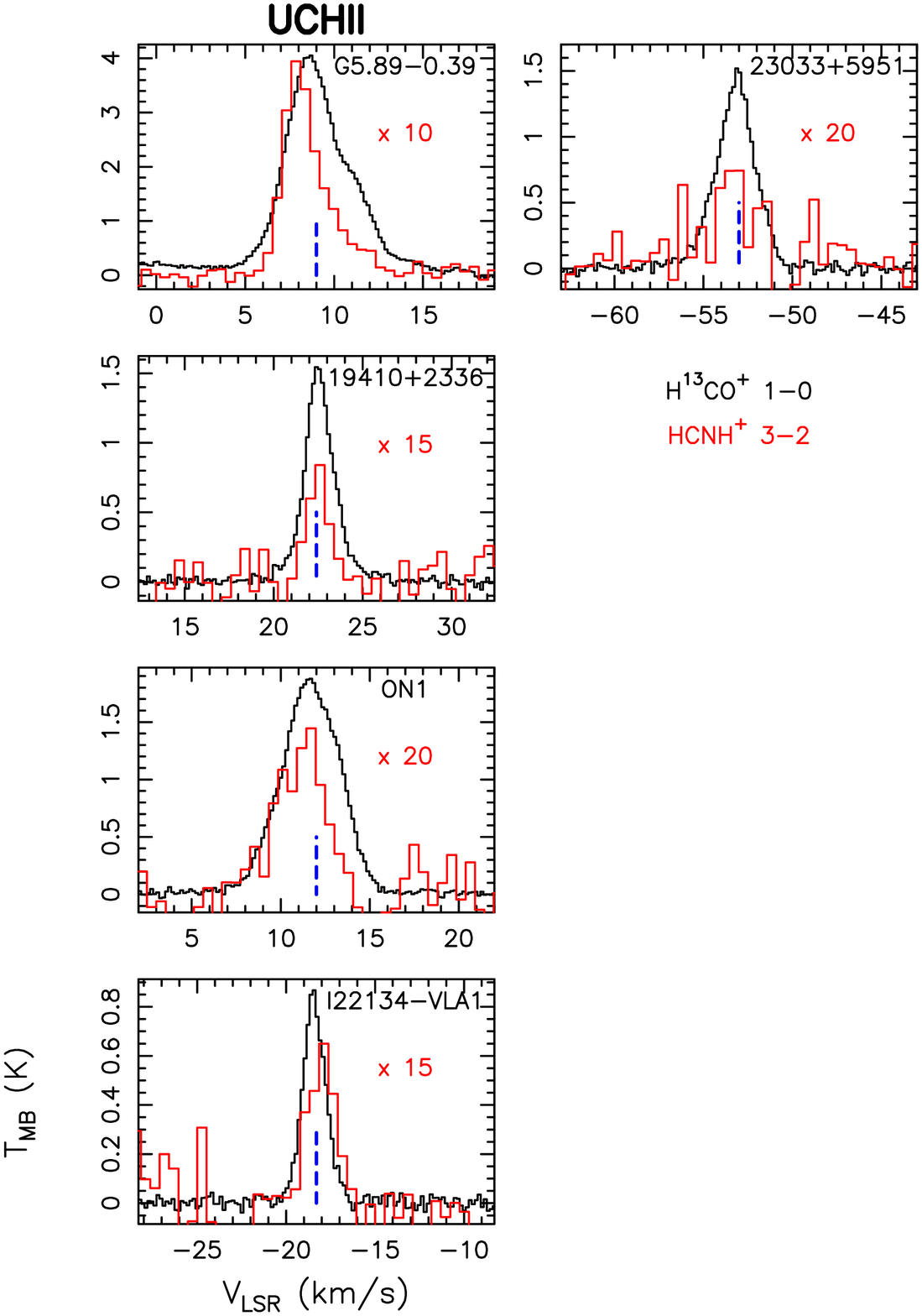}}
\caption{Comparison among the observed profiles of H$^{13}$CO$^+$ $J=1-0$ (black histograms) and \HCNHp\ $J=3-2$
(red histograms) lines towards, from left to right, the HMSCs, the HMPOs, and the UCHIIs detected in \HCNHp\ 
(Table~\ref{tab:results}).
The red number in each plot indicates the multiplicative factor applied to the \HCNHp\ spectrum. The blue
dashed vertical line indicates the systemic velocity used to centre the observed spectra (see e.g.
Fontani et al.~\citeyear{fontani15a}). The peak velocities of \HCNHp\ are reported in Table~\ref{tab:results}.}
\label{fig:profiles}
\end{figure*}

\section{Sample and observations}
\label{obs}

The 26 targets are listed in Table~\ref{tab:results}. They are taken from the sample presented first by
Fontani et al.~(\citeyear{fontani11}) and divided into the three gross evolutionary categories discussed in
Sect.~\ref{intro}: 10 HMSCs, 9 HMPOs, and 7 UC HIIs (see Fontani et al.~\citeyear{fontani11} for details on 
the source selection criteria).
The spectra analysed in this work are part of the dataset published in Fontani et al.~(\citeyear{fontani15a}) 
obtained with the IRAM-30m telescope. In particular, the analysed transition, \HCNHp $J = 3-2$, was 
detected in the band at 1.2~mm of that dataset covering frequencies in the range 216.0--223.78~GHz.
The spectra were observed with
a telescope beam of $\sim 11$\asec, and a spectral resolution of $\sim 0.26$ \kms. Details of the
observations (weather conditions, observing technique, calibration, pointing and focus checks) 
are given in Fontani et al.~(\citeyear{fontani15a}). The analysed line has
a rest frequency of $\sim 222329.277$~MHz, energy of the upper level $E_{\rm u}\sim 21.3$ K, {\rm Einstein 
coefficient $A_{\rm ij}= 4.61 \times 10^{-6}$ s$^{-1}$}, and degeneracy of the rotational upper level 
$g_{\rm u}=7$. All these spectral parameters are taken from the Cologne Database for Molecular
Spectroscopy (CDMS, Endres et al.~\citeyear{endres16}. The spectroscopy is determined in 
Araki et al.~\citeyear{araki98}. The dipole moment is taken from Botschwina~\citeyear{botschwina86}).
According to the collisional coefficients given in Nkem et al.~(\citeyear{nkem14}), which are $\sim 1 - 1.5\times 10^{-10}$
s$^{-1}$ in the temperature range 10--100 K, the critical density of the transition is $\sim 4 \times 10^{4}$~\cmc.
The lines were fitted with the {\sc class} package of the {\sc gildas}\footnote{https://www.iram.fr/IRAMFR/GILDAS/} 
software using standard procedures. 

\section{Results}
\label{res}

\subsection{Detection rate and line shapes}
\label{detection}

We detected \HCNHp $J=3-2$ with signal to noise ratio, S/N, higher than $\sim 3$ towards 14 out of the 26 
targets. In two cases, I20293--WC and 23033+5951, the peak main beam temperature is slightly
below the $3 \sigma$ rms, but the line profile suggests a real line at the limit of the detection
level. Hence, we have considered them as tentative detections, which provides a total of 16 detected sources
including the tentative ones (detection rate $\sim 62\%$). The transition has hyperfine structure that, however, 
cannot be resolved in our spectra because the separation in velocity of the components is smaller than the 
velocity resolution. Therefore, we fitted the lines with single Gaussian profiles. This method gives good results 
with residuals lower than, or comparable to, the noise in the spectra. 

Table~\ref{tab:results} reports the results of the Gaussian fits to the lines: integrated line intensity 
($\int T_{\rm MB}{\rm d}V$), velocity at line peak ($V_{\rm p}$), line full width at half maximum (FWHM), 
peak main beam temperature ($T_{\rm MB}^{\rm p}$), and $1 \sigma$ rms in the spectrum (1$\sigma$).
The uncertainties on $V_{\rm p}$ and on FWHM are given by the fit procedure. 
The uncertainties on $\int T_{\rm MB}{\rm d}V$ are the sum in quadrature of the error given by the fit and
the calibration error on $T_{\rm MB}$ (assumed to be $10\%$). The latter is dominant in the sum in quadrature.

The spectra and their fits are shown in Fig.~\ref{fig:spectra}. In two sources, the line profile cannot be fitted with 
a single Gaussian, i.e. in 18089--1732 and G5.89--0.39. Both spectra show an excess emission in the red tail of 
the line (see Fig.~\ref{fig:spectra}). Towards G5.89--0.39, this excess has the form of a high-velocity wing. Hence, 
it likely arises from the red-shifted lobe of the outflow associated with the embedded UC HII region 
(Zapata et al.~\citeyear{zapata20}). 
Towards 18089--1732, the excess emission has the shape of a Gaussian peak with intensity almost half of the 
main peak. Hence, in this case it is more likely due either to a secondary velocity feature, or to another line blended 
with the \HCNHp (3--2) one. High excitation lines ($E_{\rm u}\sim 200$K) of CH$_3$OCH$_3$, detected in this source 
(Coletta et al.~\citeyear{coletta20}), are predicted at $\sim 222326$ MHz, which could hence 
contribute to the secondary peak centred approximately at $\sim 222327$ MHz. Moreover, the profile
of the H$^{13}$CO$^+$ $J=1-0$ does not show this secondary peak (Fig.~\ref{fig:profiles}), and hence 
a second velocity feature is very unlikely.
A contribution from the red lobe of the outflow associated with the embedded Hyper-compact HII region is also 
possible (Beuther et al.~\citeyear{beuther10}), but because, again, it is not seen in H$^{13}$CO$^+$ $J=1-0$, 
its contribution to this secondary peak should be negligible.
In general the lack of high-velocity wings, except maybe in the two cases discussed
above, indicates that the emission of the line does not arise from shocks and/or outflows,
but likely from more quiescent material.
 
Because \HCNHp\ was detected in a handful of star-forming regions before this study, it is
not yet clear which kind of material (i.e. with which physical and kinematical properties) is responsible 
for the emission of this molecule. To investigate this, in Fig.~\ref{fig:profiles} we compare the profiles of 
\HCNHp $J=3-2$  with those of \HCOpI $J=1-0$ analysed in \citet{fontani18}, thought to be a good tracer of the envelope of 
star-forming cores. Inspection of the plot suggests that in some sources (e.g. G028--C1, AFGL5142--EC, 05358--mm1,
I20293--MM1, 19410+2336), the two lines have similar peak velocity, suggesting a common origin in the envelope of 
the cores. On the other hand, in some targets, e.g. G034--F1, I20293--WC, I22134--VLA1, and G5.89-0.39, the 
\HCNHp\ emission does not coincide with the central velocity of \HCOpI, even though towards G034--F1 and 
I20293--WC, the H$^{13}$CO$^+$ $J=1-0$ shows multiple velocity features, and the \HCNHp $J=3-2$ emission 
is centred on one of these. These second velocity features, in principle, could also be originated by 
self-absorption or polluting lines from other species. These scenarios, however, are both unlikely in 
G034--F1 and I20293--WC. In fact, self-absorption signatures were never observed in other lines towards
these sources, and polluting lines are very unlikely in cold HMSCs (like these two targets). Furthermore, a
second velocity feature towards G034--F1 was already found in the HCN isotopologues by \citet{colzi18a}.

About the line widths, the stronger H$^{13}$CO$^+$ lines are generally broader than the \HCNHp\ ones, except 
than in a few targets (G034--G2, 18089--1732, I22134--VLA1). Also, we see no hints of high velocity wings
in \HCNHp\ $J=3-2$, except than in the already discussed spectrum of G5.89--0.39. However, high velocity 
wings could not be detected due to the limited signal-to-noise ratio in the spectra. All this indicates that:
(1) the gas emitting \HCNHp\ is mostly quiescent, or not clearly associated with shocks, and (2) there is not a 
general agreement with the profiles of the \HCOpI $J=1-0$ lines. We propose that in the targets in which both 
lines have the same velocity peak, the \HCNHp\ $J=3-2$ line arises from the envelope of the cores, but likely
from a more compact portion of that traced by \HCOpI $J=1-0$. 
In fact, the \HCNHp\ $J=3-2$ line has a higher energy of the upper level than \HCOpI\ $J=1-0$ (21.3~K against
4~K), and the spectra were observed with different beam sizes ($\sim 11$\asec\ against $\sim 28$\asec).

\subsection{Total column densities and abundances of \HCNHp}
\label{coldens}

Assuming that the lines are optically thin and in Local Thermodynamic Equilibrium (LTE), 
from $\int T_{\rm MB}{\rm d}V$ we have computed the beam-averaged total 
column densities of \HCNHp, \Ntot (\HCNHp), using Eq. (1) of \citet{fontani18}. For undetected sources, we have 
estimated upper limits on \Ntot (\HCNHp) from the upper limit on the integrated intensity. This was computed 
from the relation $\int T_{\rm MB}d{\rm v} = 3\sigma\frac{\sqrt{\pi}}{2\sqrt{{\rm ln2}}}{\rm FWHM}$,
which expresses the integral in velocity of a Gaussian line with peak intensity given by the $3\sigma$ rms 
in the spectrum. We assumed as FWHM the average value obtained from the detected lines 
in each evolutionary group which the undetected source belongs to: 2.5~\kms\ for HMSCs, 1.6~\kms\ for HMPOs, 
2.4~\kms\ for UCHIIs (see Table~\ref{tab:results}).

The assumption of optically thin emission is consistent with the low abundance of the molecule and with the fact
that the observed line shapes have generally no hints of high optical depths such as, e.g., asymmetric or flat topped 
profiles. Only towards G034--G2, the almost flat-top line shape could indicate non-negligible optical depths. Hence,
in this source, the \HCNHp\ total column density should be regarded as a lower limit.
The assumption of LTE is also reasonable because the critical density of the line ($\sim 4 \times 10^{4}$~\cmc, see
Sect.~\ref{obs}) is smaller than, or comparable to, the average H$_2$ volume densities of the sources.
These, calculated within a beam of 28\asec, are in between $10^{4}$ and $10^{6}$~\cmc\ 
(see e.g.~Fontani et al.~\citeyear{fontani18} and Sect.~\ref{discu}). A few sources have 
H$_2$ volume density of $\sim 1-2 \times 10^{4}$~\cmc, i.e. the HMSCs G034--F2, G034--F1, and G028--C1, 
and the UCHII I22134--VLA1 (see Sect.~\ref{discu}). For these targets, the approximation could not be valid.
We have estimated by how much \Ntot (\HCNHp) would change in the conservative case in which 
the excitation temperature is lower than the kinetic temperature by 10~K: \Ntot (\HCNHp) would increase 
by a factor $\sim 4$. Therefore, in the following the column densities of these targets should be considered 
lower limits.

The total column densities, averaged within the telescope beam of 11\asec, are in the range $0.5-10 \times 10^{13}$\cmq.
If we analyse separately the three evolutionary groups, we find hints of possible differences. In fact, the average 
values are: $\sim 1.5 \times 10^{13}$\cmq\ for the HMSCs, $\sim 1.25 \times 10^{13}$\cmq\ for the HMPOs, 
$\sim 2.7 \times 10^{13}$\cmq\ for the UCHIIs. However, the average value in the UCHIIs is strongly
biased by G5.89--0.39, which is by far the most luminous object in the sample. Without G5.89--0.39, the 
average value in the UCHIIs is $\sim 1.3 \times 10^{13}$\cmq, consistent with those of the other two 
evolutionary groups.

We have computed abundances of \HCNHp\ relative to H$_2$ by using the H$_2$ column densities
published in \citet{fontani18}. These were average values derived in an angular diameter of 28\asec, 
hence the \HCNHp\ column densities have been smoothed to this larger angular diameter.
The corresponding abundances, [\HCNHp], are listed in Table~\ref{tab:results}, and range from $\sim 0.9$ to 
$\sim 14\times 10^{-11}$. Interestingly, HMPOs and UCHIIs have average [\HCNHp] of the order of
$10^{-11}$, and consistent among them (specifically, $\sim 2.5$ and $\sim 3.7 \times 10^{-11}$, respectively), 
while HMCSs have an average [\HCNHp] of $\sim 7.3 \times 10^{-11}$, i.e. a factor 2--3 higher. In particular, the
HMSCs with the highest abundances are all classified as "cold", or "quiescent", by \citet{fontani11}, and
have [\HCNHp] of the order of $\times 10^{-10}$. 

\begin{figure*}
{\includegraphics[width=6cm]{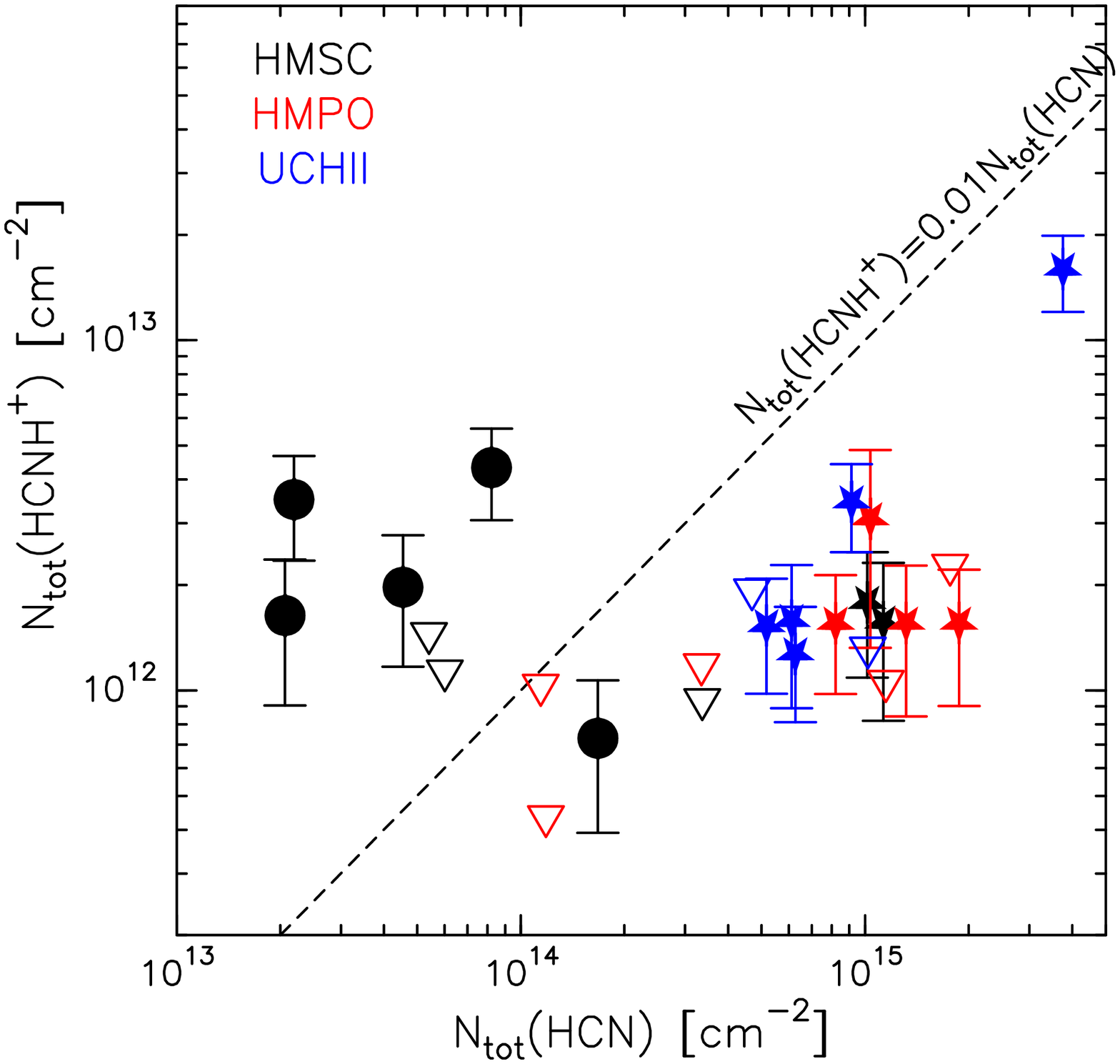}
\includegraphics[width=5.8cm]{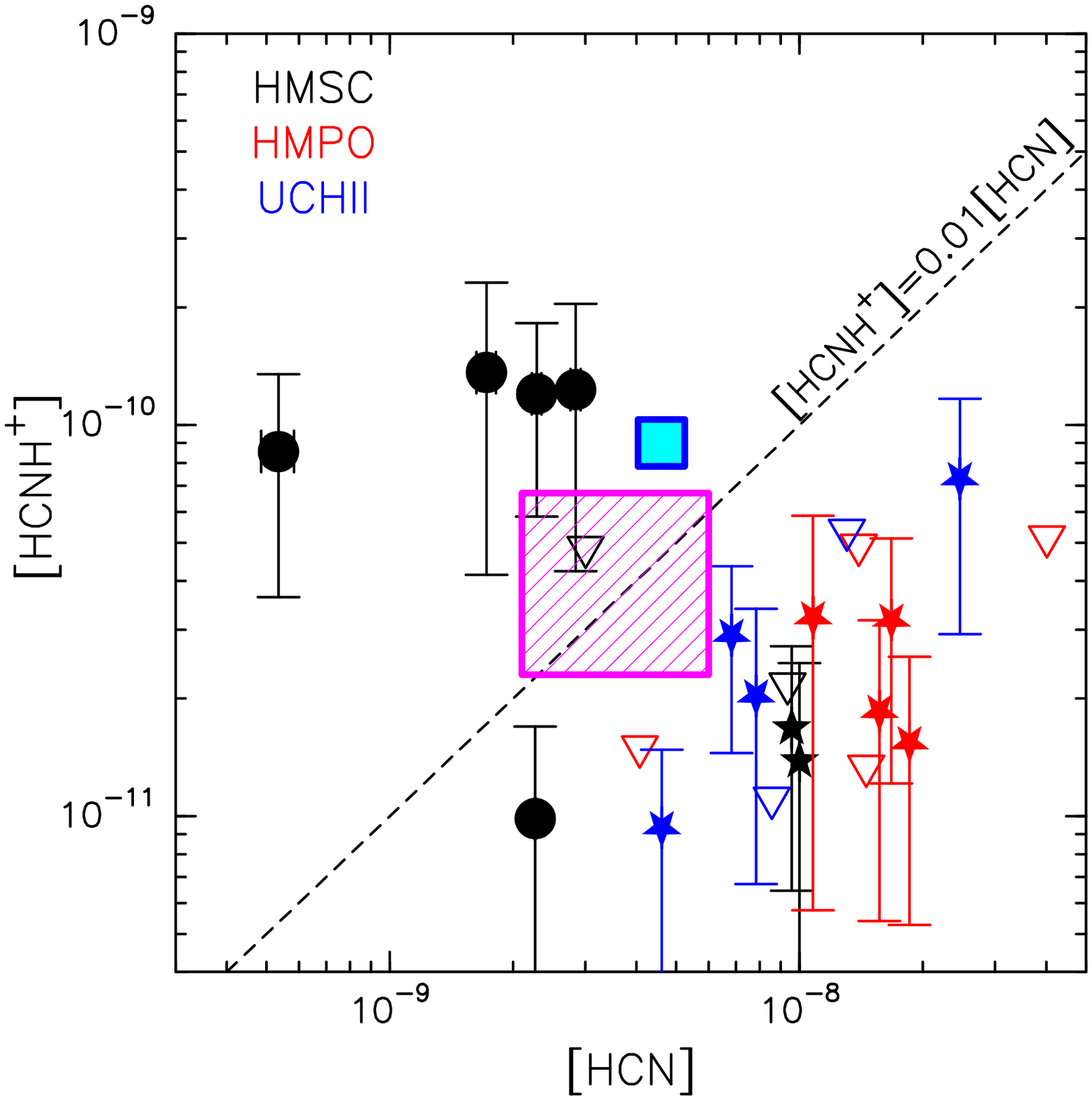}
\includegraphics[width=6cm]{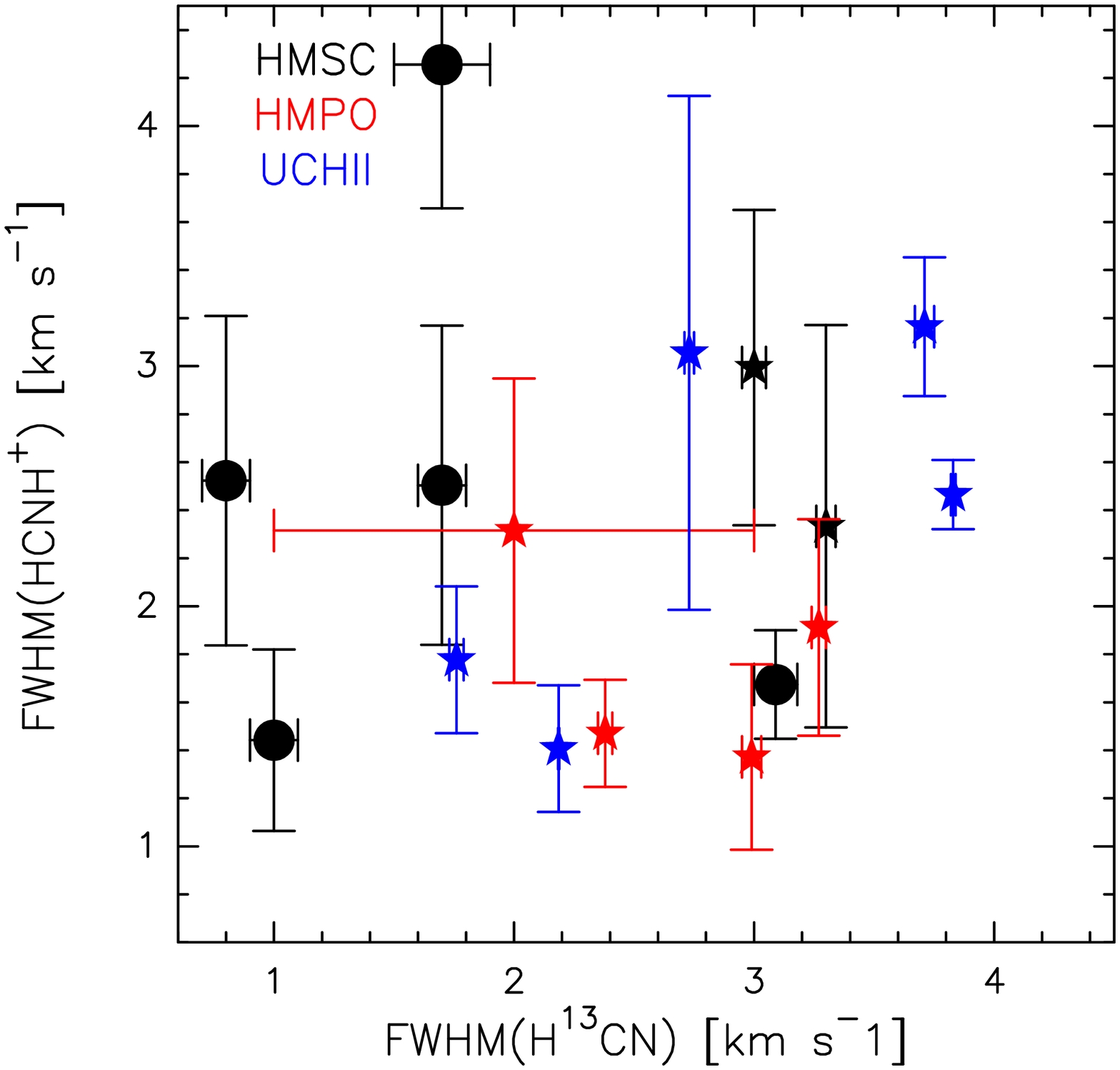}}
      \caption{{\it Left panel}: comparison between \Ntot (\HCNHp) and \Ntot (HCN). The latter was
      derived from the \H13CN $J=1-0$ line (Colzi et al.~\citeyear{colzi18b}), assuming the same \Tex,
      by correcting it for the $^{12}$C/$^{13}$C ratio obtained at the Galactocentric distance of the 
      sources from the trend of~\citet{yan19}. \Ntot (\HCNHp)
      has been rescaled to the (larger) beam of the \H13CN\ observations, i.e. $\sim 28$\asec.
      Red stars are HMPOs; blue stars are UCHIIs; black symbols are HMSCs, among which 
      the filled circles represent the "quiescent" ones
      while the stars indicate the "warm" ones (See Table~\ref{tab:results}). Triangles indicate
      upper limits on \Ntot (\HCNHp). The dashed line is the locus where \Ntot (\HCNHp) = 0.01 \Ntot (HCN).
      \newline
      {\it Middle panel}: same as left panel, but for the fractional abundances with respect to H$_2$ calculated
      on a beam of $\sim 28$\asec.
      The purple rectangle indicates the range of abundances that reproduces the observed [HCN], 
      and [\HCNHp] in the predictions of the Warm Model (WM, see Sect.~\ref{model} and right panel
      of Fig.~\ref{fig-abu}).
 Similarly, the cyan square indicates the prediction of the Cold Model (CM, see Sect.~\ref{model} and left panel
      of Fig.~\ref{fig-abu}). 
      \newline
       {\it Right panel}: comparison between the line FWHM of \HCNHp, derived in this work from the $J=3-2$ line,
       and \H13CN, derived by \citet{colzi18a} from the $J=1-0$ line at $\sim 87$~GHz. 
       The symbols have the same meaning as those in the left and middle panels.}
         \label{fig:coldens}
\end{figure*}

\begin{figure*}
{\includegraphics[width=6cm]{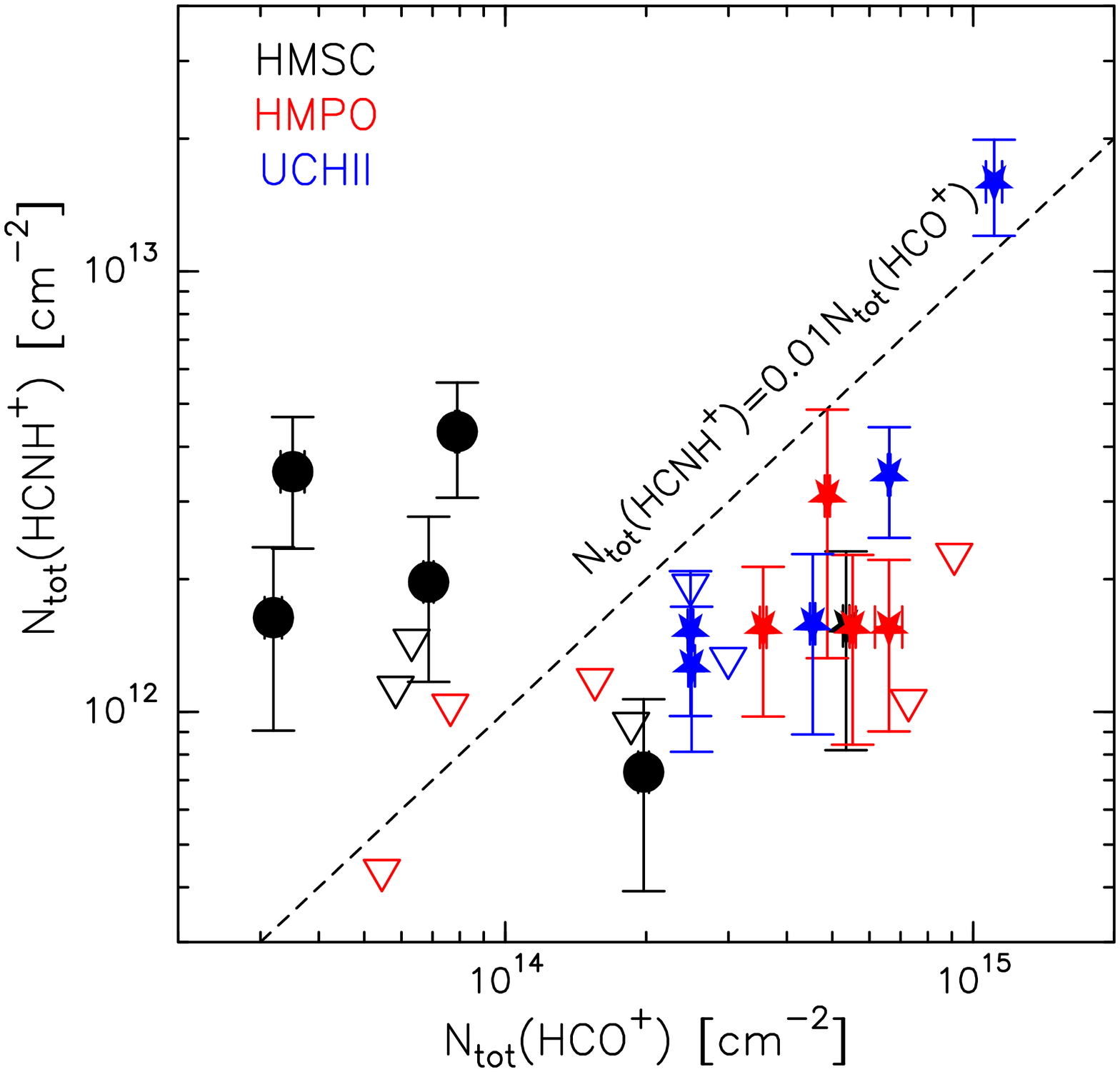}
\includegraphics[width=6cm]{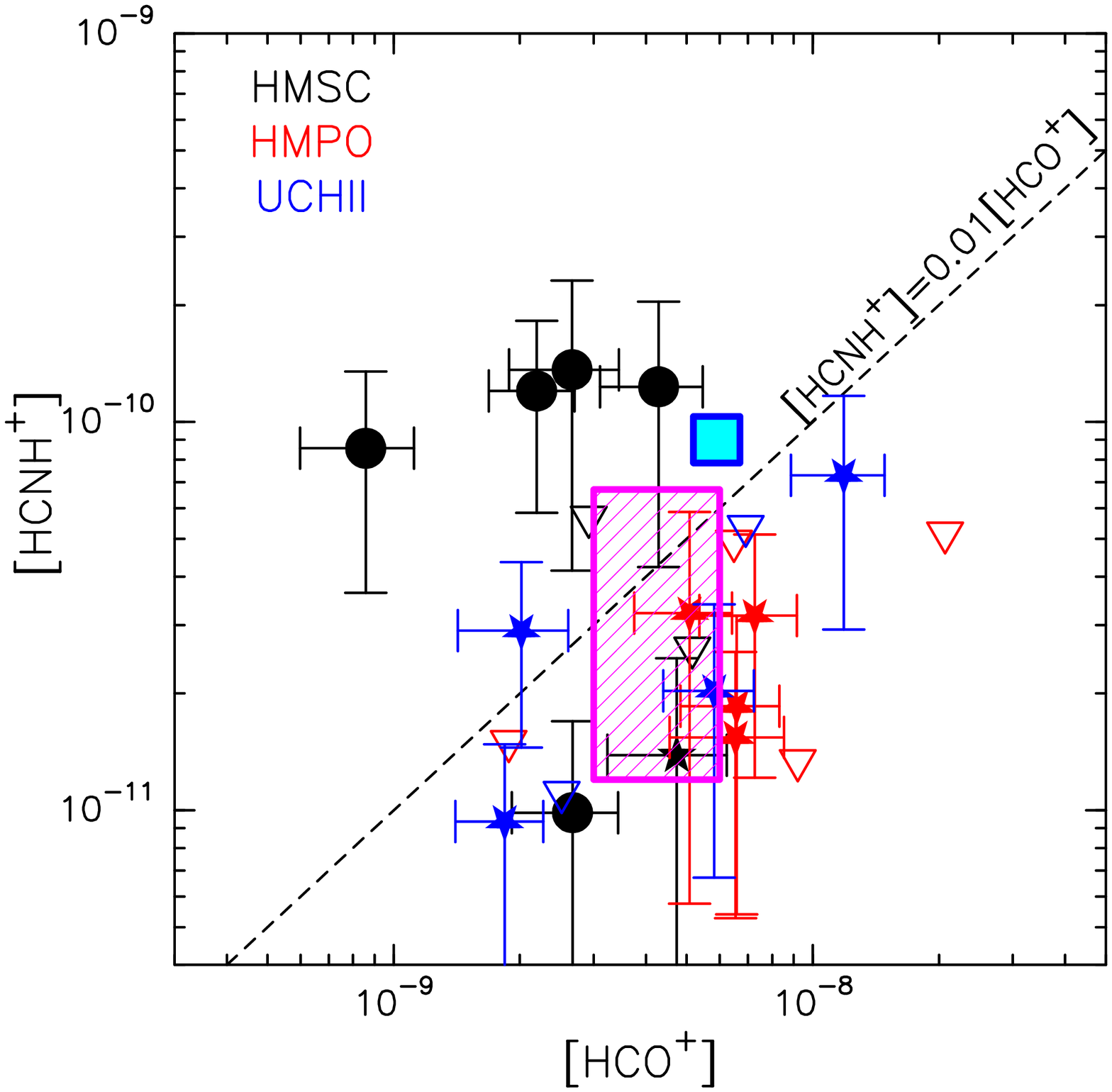}
\includegraphics[width=6cm]{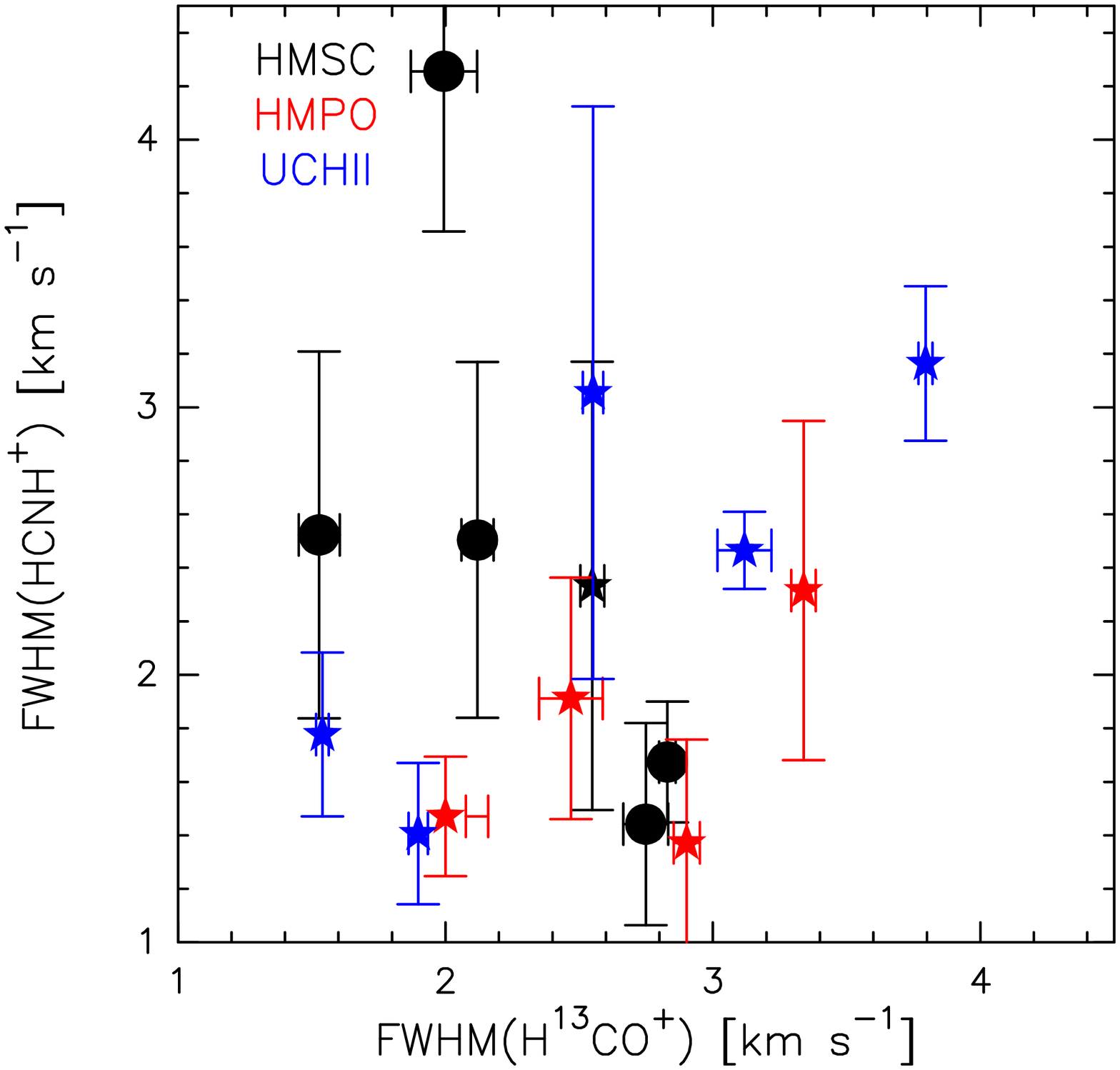}}
 \caption{{\it Left panel}: comparison between \Ntot (\HCNHp) and \Ntot (HCO$^+$).
The latter has been computed from \Ntot (H$^{13}$CO$^+$) taken from \citet{fontani18},
by correcting it for the $^{12}$C/$^{13}$C ratio obtained at the Galactocentric distance 
of the sources from the trend of~\citet{yan19}.
 \newline
 {\it Middle panel}: same as left panel, but for the abundances relative to H$_2$. As in middle panel
 of Fig.~\ref{fig:coldens}, the purple rectangle and the cyan square show the predictions of
 the Warm and Cold Models (Sect.~\ref{model} and Fig.~\ref{fig-abu}), respectively, that best reproduce 
 the observed abundances of the three species \HCNHp, HCN, and HCO$^+$.
 \newline
 {\it Right panel}: comparison between the line FWHM of \HCNHp, derived in this work from the $J=3-2$ line,
       and H$^{13}$CO$^+$, derived in Fontani et al.~(\citeyear{fontani18}) from the $J=1-0$ line at $\sim 87$~GHz. 
       The symbols have the same meaning as those in the left and middle panels.}
          \label{fig:coldens-HCO+}
\end{figure*}


\section{Discussion of the observational results}
\label{discu}

One of the most important and direct findings from the parameters in Table~\ref{tab:results}
is that the HMSCs classified as quiescent have \HCNHp\ abundances larger by about an order 
of magnitude with respect to warm(er) and/or more evolved objects. 
This suggests that in cold and quiescent cores some formation pathways of \HCNHp\ are 
more efficient than in warmer cores, and/or that some destruction pathways are less efficient.
The chemistry of \HCNHp\ is thought to be related to that of HCN, \HCOp, and CN, as
discussed in Qu\'enard et al.~(\citeyear{quenard17}). However, Qu\'enard et al.~(\citeyear{quenard17}) 
focussed on the pre-stellar core L1544, i.e. a cold core with a known physical structure. 
Which are the dominant formation pathways in warmer environments (and under different 
physical conditions in general), are not yet clear.

%
%
%
%
%
%

To investigate the chemical origin of \HCNHp, we have searched for correlations between some 
physical parameters of  \HCNHp\ and those of HCN, \HCOp, and CN. In Fig.~\ref{fig:coldens} we 
compare column densities, abundances, and FWHM of \HCNHp\ and HCN. 
Fig.~\ref{fig:coldens-HCO+} shows the same comparison between \HCNHp\ and
\HCOp. Note that the column densities of HCN and \HCOp\ have been calculated from
the data of H$^{13}$CN and H$^{13}$CO$^+$ published in Colzi et al.~(\citeyear{colzi18a}) 
and Fontani et al.~(\citeyear{fontani18}), respectively.
The column densities of both \H13CN\ and H$^{13}$CO$^+$ were computed assuming the 
\Tex\ in Table~\ref{tab:results}, and converted to those of the main isotopologues using the 
$^{12}$C/$^{13}$C ratio at the Galactocentric distance of the sources. This latter was
derived from the most recent trend of~\citet{yan19}. Calculated $^{12}$C/$^{13}$C ratios
are in the range $\sim 40 - 70$. The uncertainties on these ratios, calculated propagating the 
errors on the parameters of the Galactocentric trend, are of the order of the 30 $\%$.
As for \HCNHp, the final \Ntot (HCN) and \Ntot (HCO$^+$) were scaled to the same beam 
of $\sim 28$\asec, and the abundances [HCN] and [HCO$^+$] were computed 
dividing \Ntot (HCN) and \Ntot (HCO$^+$) by the same \Ntot (H$_2$) used for \HCNHp.

Two relevant results emerge clearly from the plots in Figs.~\ref{fig:coldens} and ~\ref{fig:coldens-HCO+}: 

\begin{itemize}
\item[(i)] the abundance ratios [\HCNHp]/[HCN] and [\HCNHp]/[\HCOp] are both $\leq 0.01$ for all objects 
except for the four quiescent HMSCs G034--G2, G034--F2, G034--F1, and G028--C1, for which they are one 
order of magnitude above this threshold. This difference could be even more pronounced because, as 
discussed in Sect.~\ref{coldens}, for three of these four sources [\HCNHp] could be underestimated
because the transition could be sub-thermally excited;
\item[(ii)] the FWHM of the lines of different species do not appear clearly correlated. However, if
one excludes the four quiescent HMSCs mentioned in item (i), a tentative correlation is apparent in 
both Fig.~\ref{fig:coldens} and Fig.~\ref{fig:coldens-HCO+}, with Spearman $\rho$ correlation coefficients 
$\sim 0.55$ and $\sim 0.63$, respectively.
\end{itemize}

These findings suggest that the dominant formation pathways of \HCNHp\ in cold and warm regions are 
likely different. Only one quiescent HMSC, I20293--WC, does not follow this different behaviour. 
However, the detection of \HCNHp\ is tentative towards this source (see Table~\ref{tab:results}), 
and the core, even though considered quiescent based on its low kinetic temperature (Fontani et al.~\citeyear{fontani11}), 
is embedded in a star-forming region harbouring a more evolved object (e.g. core 
I20293--MM1, included in the HMPOs). Therefore, the chemistry in this target could be
influenced by the complex environment in which it is embedded.

\begin{figure*}
{\includegraphics[width=5.9cm]{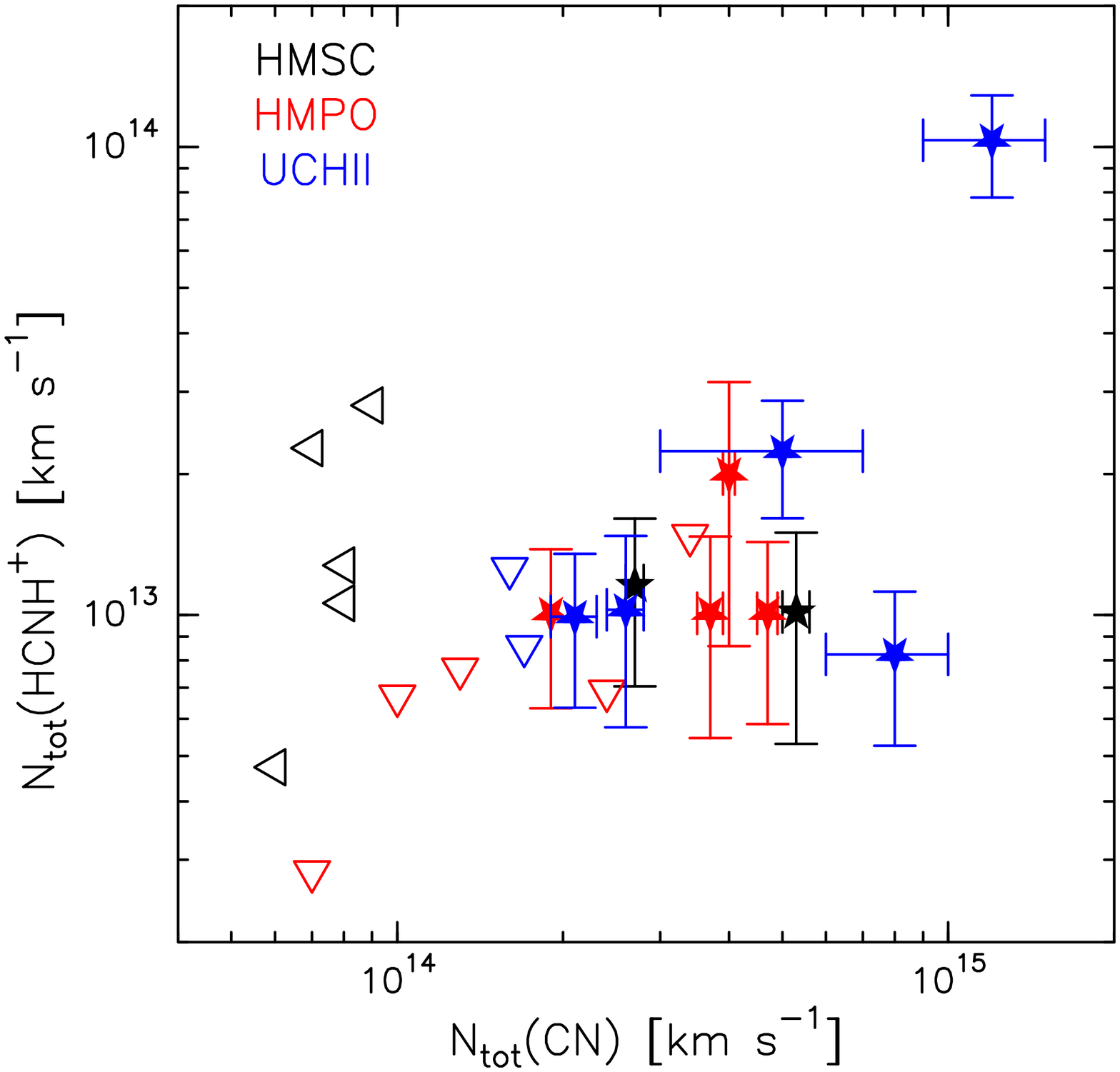}
\includegraphics[width=5.9cm]{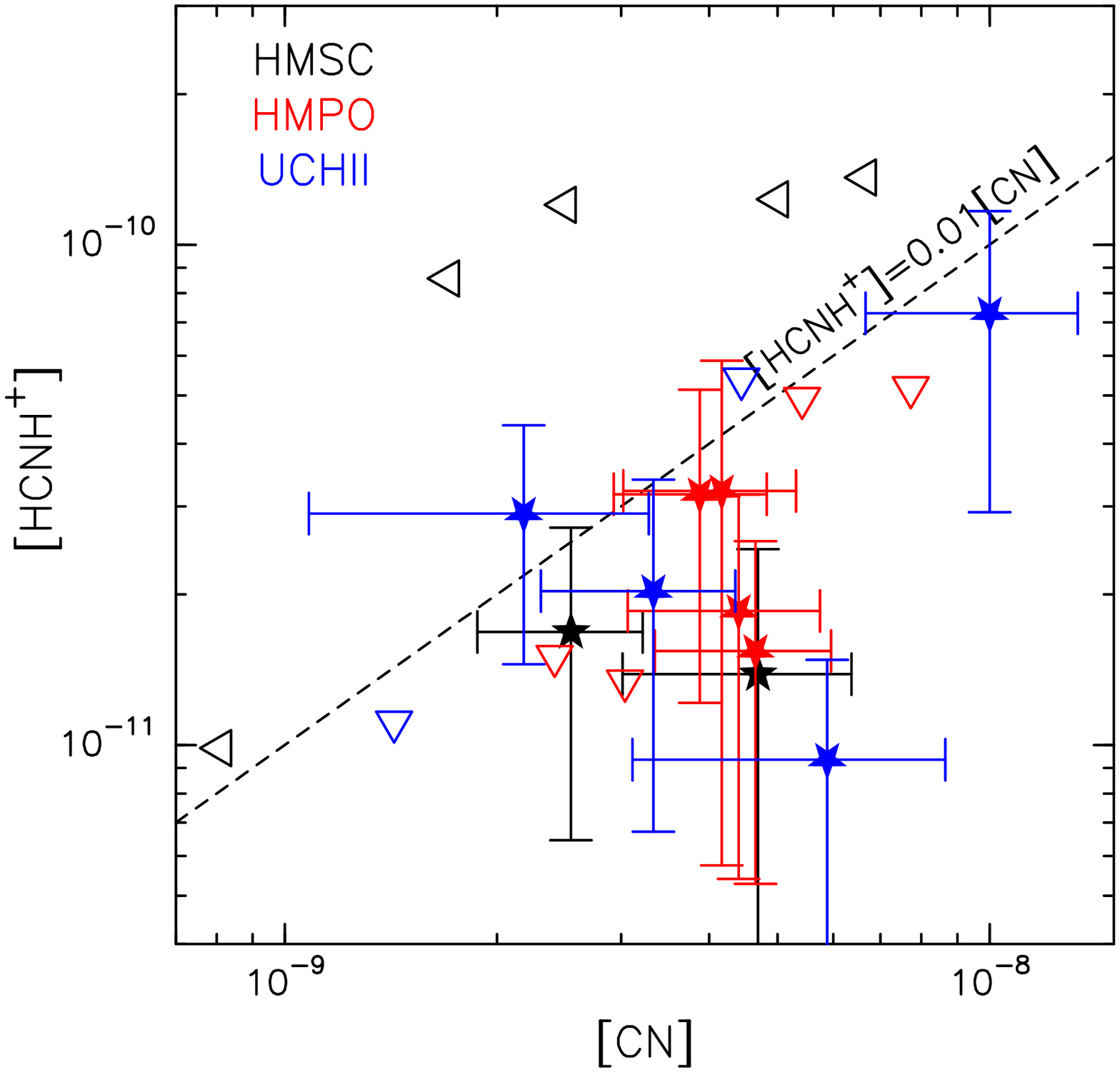}
\includegraphics[width=5.9cm]{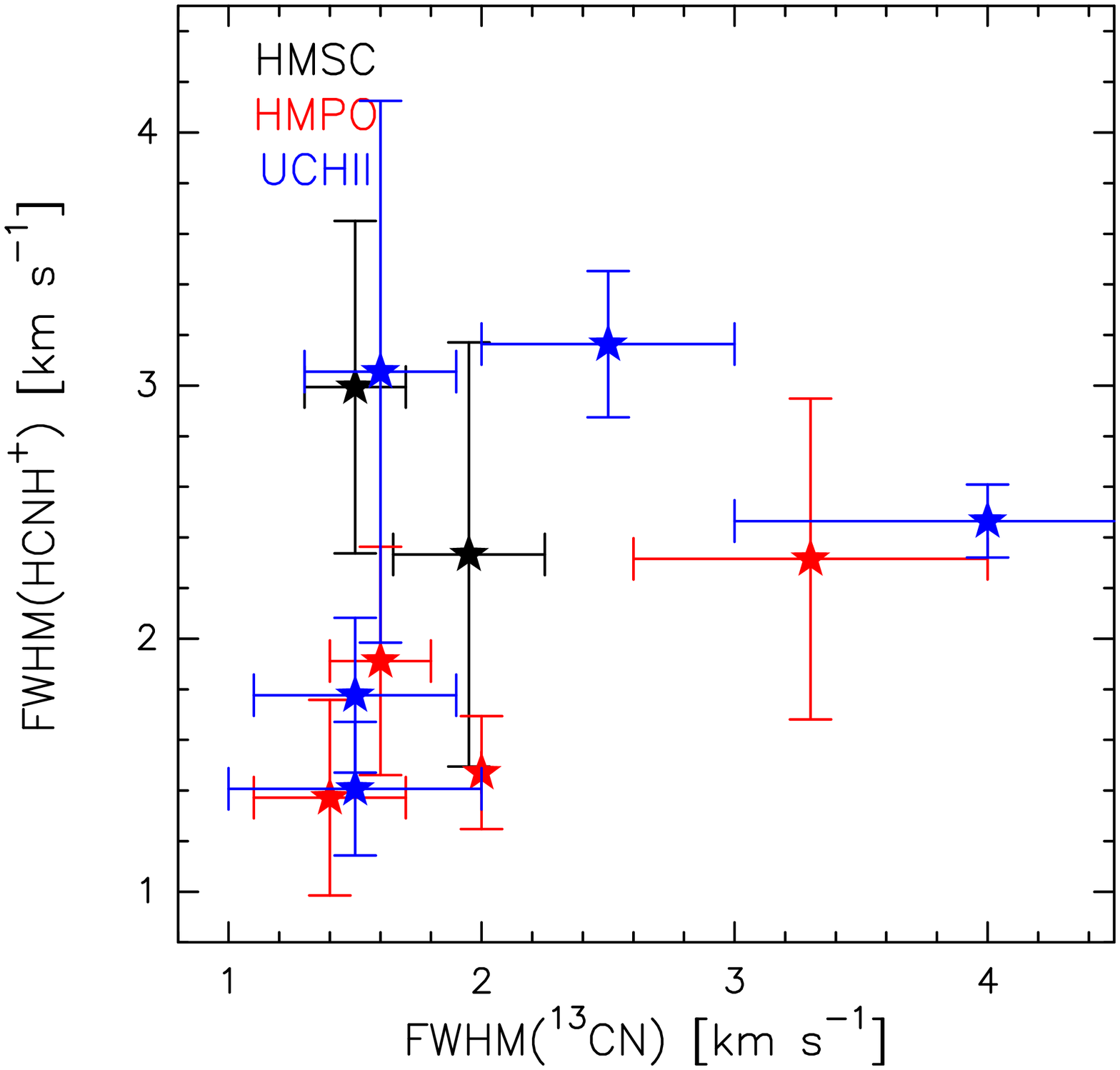}}
      \caption{Same as Fig.~\ref{fig:coldens} for the comparison between \HCNHp\ and CN.
      For CN, \Ntot\ (left panel), [CN] (middle panel), and FWHM (right panel) are derived from observations of the 
      $^{13}$CN $N=2-1$ line at $\sim 217$~GHz, published in \citet{fontani15b}.
      \Ntot (\HCNHp) and \Ntot (CN) are averaged within the same beam of 
      $\sim 11$\asec, and triangles with vertex pointing to the left side of the plot are upper limits
      on \Ntot (CN), while those pointing to the bottom are upper limits on \Ntot (\HCNHp).
      [\HCNHp] and [CN] are averaged within a beam of 28\asec.
               }
         \label{fig:coldens-CN}
\end{figure*}

Fig.~\ref{fig:coldens-CN} shows the same comparison of physical parameters as in Fig.~\ref{fig:coldens}
and Fig.~\ref{fig:coldens-HCO+} for CN. 
The properties of this latter species have been derived from $^{13}$CN by \citet{fontani15b}.
Inspection of Fig.~\ref{fig:coldens-CN} shows a less clear dichotomy between HMSCs and the more
evolved targets, but unfortunately the comparison is affected by the non-detection of $^{13}$CN in
all the quiescent HMSCs. 
However, also in this case the [\HCNHp]/[CN] ratio in warm and/or evolved objects is $\leq 0.01$.
The FWHM of the lines do not seem correlated either, and hence the possible chemical relation
between \HCNHp\ and CN seems unlikely.

\begin{figure*}
{\includegraphics[width=4.4cm]{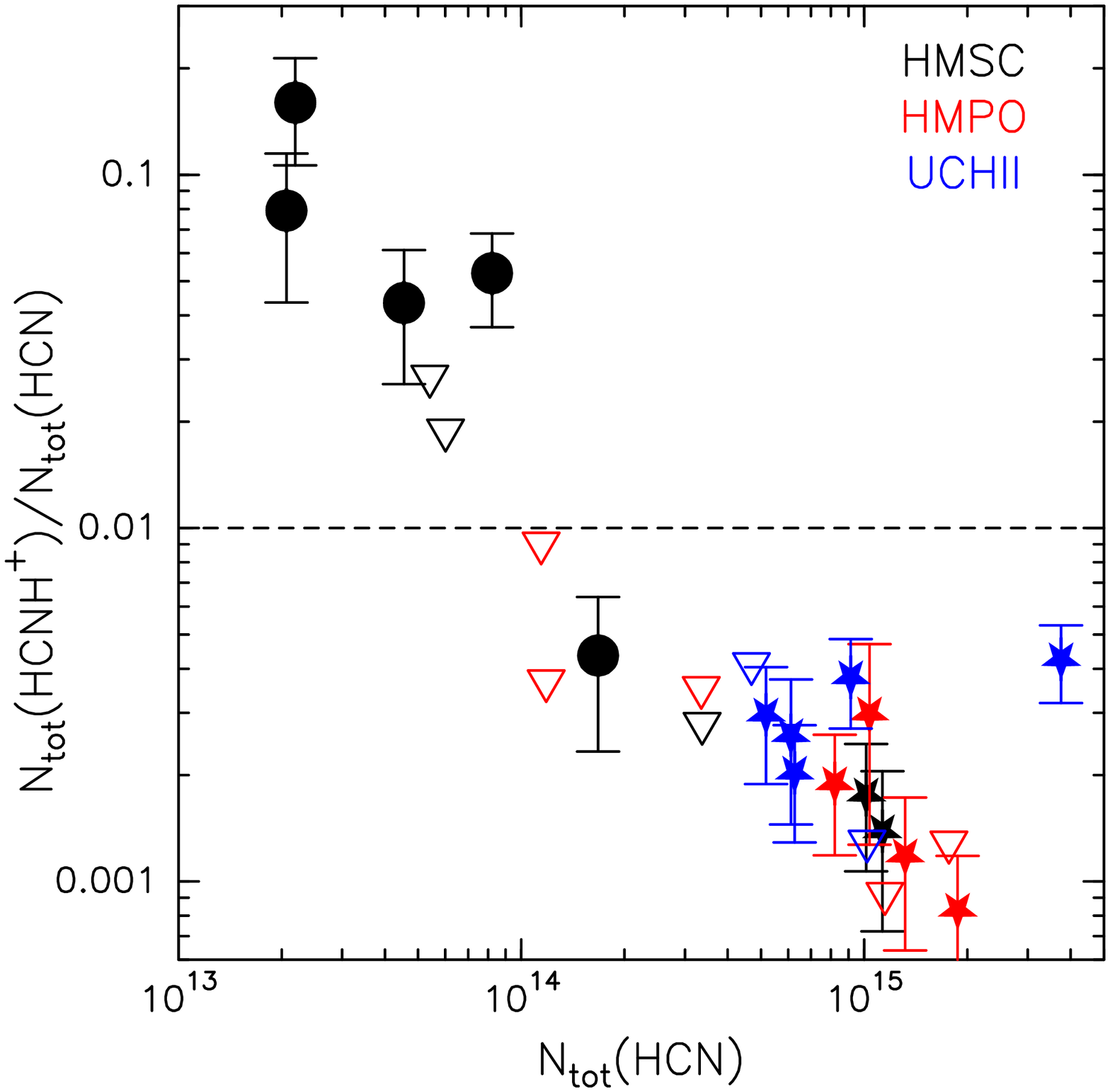}
\includegraphics[width=4.4cm]{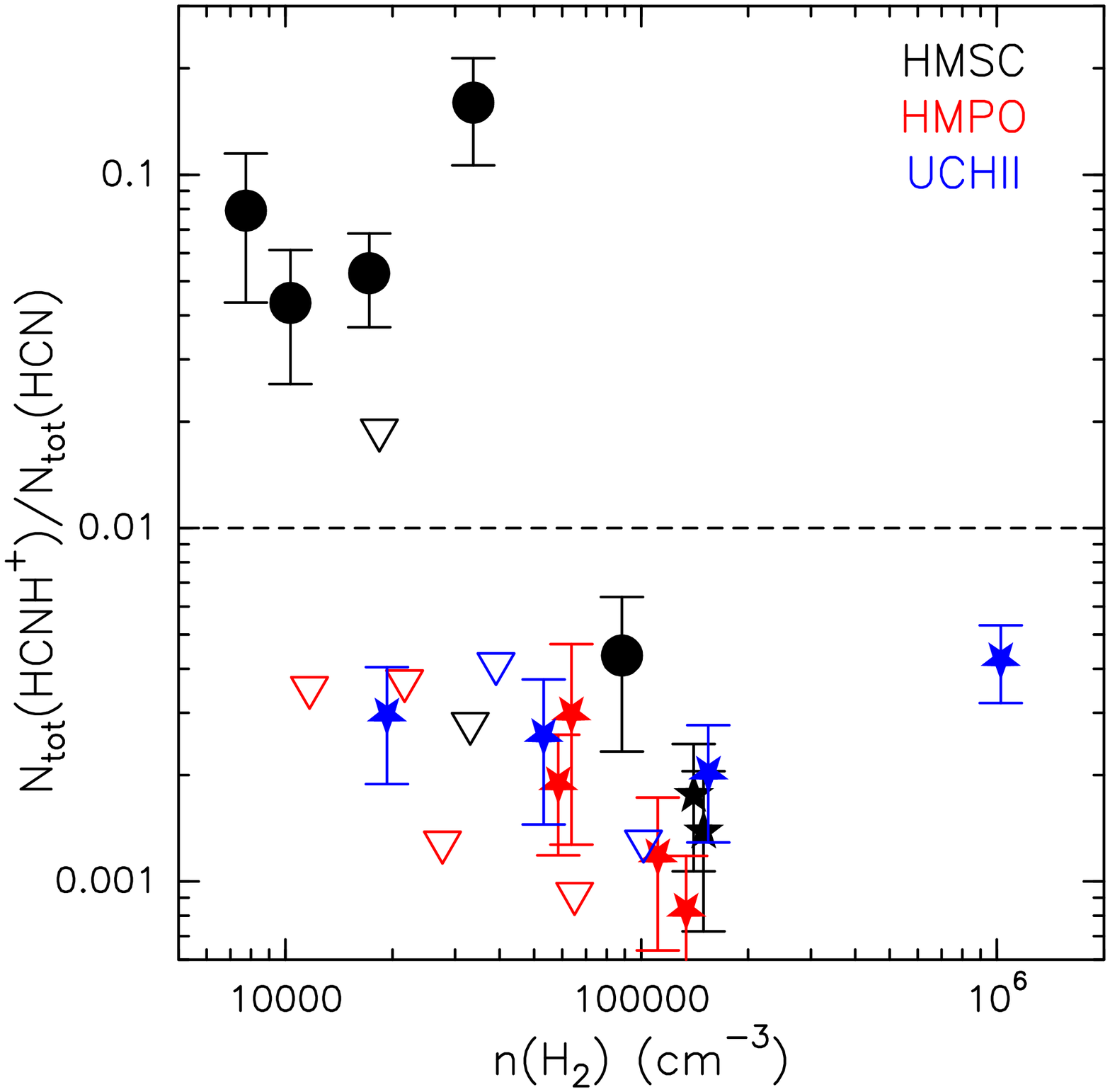}
\includegraphics[width=4.5cm]{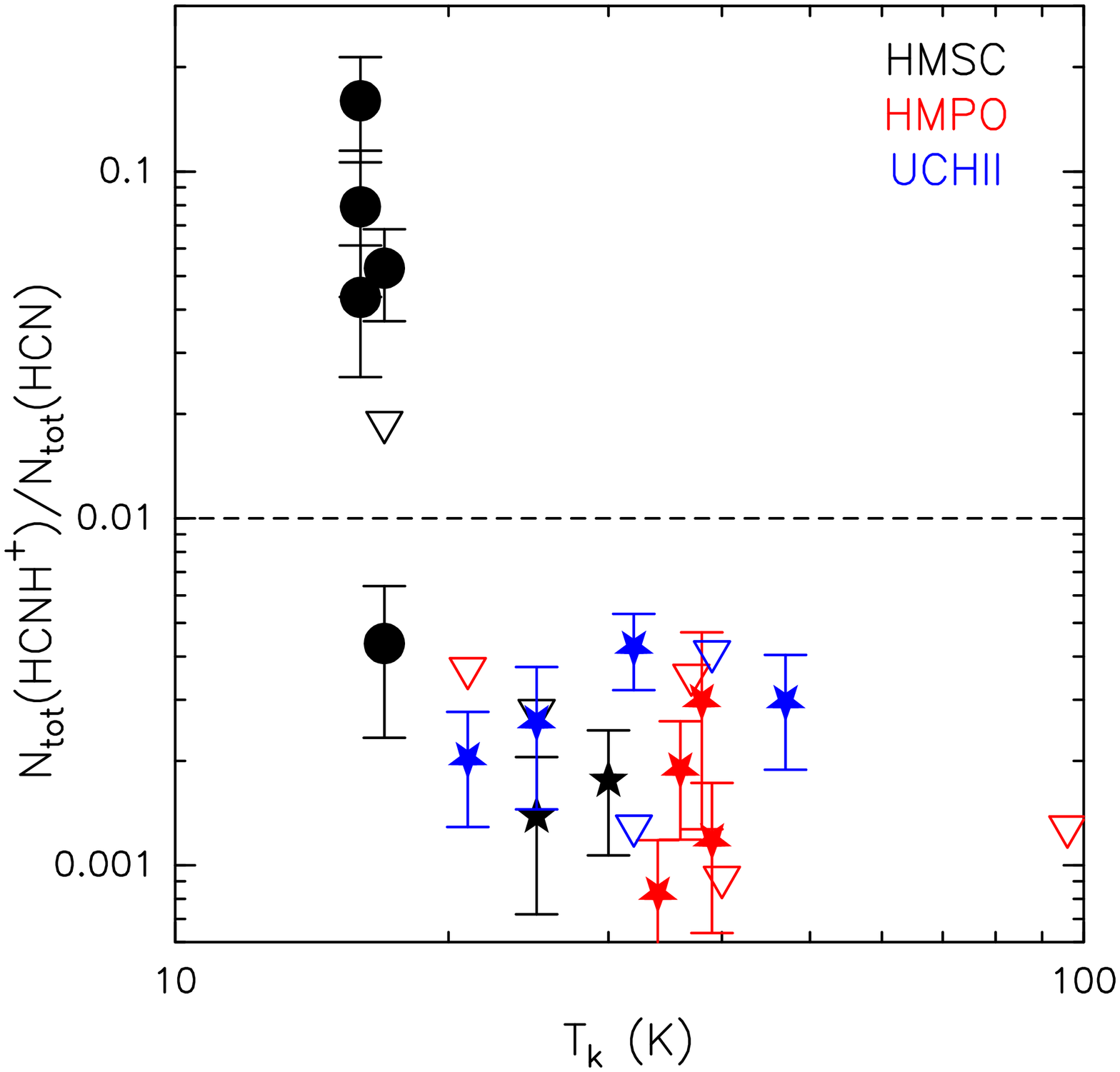}
\includegraphics[width=4.4cm]{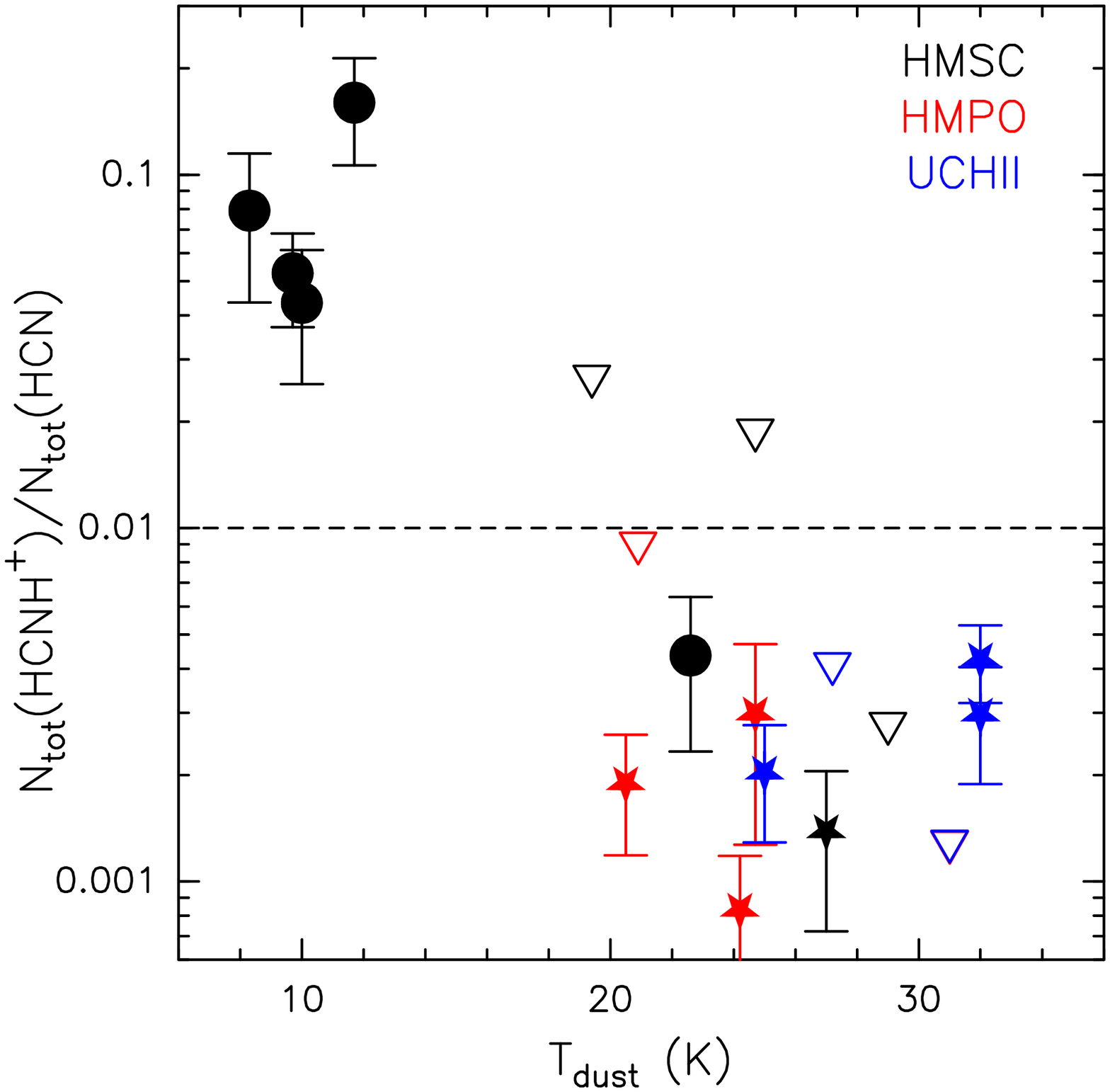}}
 \caption{Comparison between the column density ratio \Ntot (\HCNHp)/\Ntot(HCN) and the following physical
 parameters of the cores (from left to right): \Ntot (HCN), the average H$_2$ volume density $n$(H$_2$), the 
 gas kinetic temperature \Tk, and the dust temperature \Td.
 $n$(H$_2$) is an average value over an angular region of 28\asec\ (corresponding to the beam size of the \HCNHp\
 spectra), computed from the H$_2$ total column densities given in \citet{fontani18}. 
 \Tk\ is taken from \citet{fontani11} and it is used also in \citet{colzi18a} to compute the HCN column densities. 
  \Td\ was calculated by Mininni et al. (submitted) from grey-body fits to the
 Spectral Energy Distribution of the sources derived from the Herschel bands.
 The dashed line in each plot indicates \Ntot (\HCNHp)/\Ntot(HCN) = 0.01.
 }
\label{fig:rapcoldens-varie}
\end{figure*}

Finally, in Fig.~\ref{fig:rapcoldens-varie} we plot the column density ratio \Ntot (\HCNHp)/\Ntot (HCN) as a function of 
the following core physical parameters: \Ntot (HCN); volume density of H$_2$, $n$(H$_2$);
gas kinetic temperature, \Tk\ (Fontani et al.~\citeyear{fontani11});
dust temperature, \Td\ (Mininni et al, submitted).
$n$(H$_2$) is an average value within 28\asec\ computed from the column density of H$_2$ 
given in \citet{fontani18} assuming spherical sources for simplicity. Based on this plot, the different behaviour of cold and 
warm sources is very clear: colder and less dense objects have \Ntot (\HCNHp)/\Ntot (HCN) 
ratios higher than warmer and denser ones by about an order of magnitude.

\section{Chemical modelling}
\label{model}
 
In this Section we investigate the main routes of formation and destruction of HCNH$^{+}$ and HCN, together with their possible chemical relation 
with HCO$^{+}$, for two different chemical models, which represent the colder and warmer conditions of the sample of high-mass star-forming regions 
studied in this work.
 
\subsection{Description of the models}
\label{chem-model}

For the chemical simulations we use our gas-grain chemical code, recently described in Sipil\"a et al.~(\citeyear{sipila19a},~\citeyear{sipila19b}). 
The chemical networks used here contain deuterium and spin-state chemistry (see Sipil\"a et al.~\citeyear{sipila19b}, and references therein). 
While we do not explicitly consider deuterium chemistry in this paper, the inclusion of spin-state chemistry is important due 
to its effect on nitrogen chemistry (e.g.~Dislaire et al.~\citeyear{dislaire12}).
The chemical network contains a combined total of 881 species and 40515 reactions (37993 gas-phase reactions and 
2519 grain-surface reactions). We have excluded all molecules that contain more than five atoms so that the chemical 
simulations can run faster; this is justified in the present context where we study the chemistry of molecules with four 
atoms or fewer.
The initial abundances adopted in the models are summarised in Table~\ref{tab:initial}.

\begin{table}
\setlength{\tabcolsep}{6pt}
\caption{Initial abundances with respect to $n_{\rm H}$. Adapted from \citet{semenov10}.}
\centering
  \begin{tabular}{cc}
  \hline
  Species   & Initial abundance\\ 
  \hline
  H$_{2}$   & 0.5\\
  He    & 9.00$\times$10$^{-2}$\\
  C$^{+}$& 1.20$\times$10$^{-4}$\\
  N    & 7.60$\times$10$^{-5}$ \\
  O & 2.56$\times$10$^{-4}$\\
   S$^{+}$   & 8.00$\times$10$^{-8}$\\
   Si$^{+}$        &8.00$\times$10$^{-9}$ \\
   Na$^{+}$         &2.00$\times$10$^{-9}$ \\
  Mg$^{+}$         & 7.00$\times$10$^{-9}$\\
Fe$^{+}$     & 3.00$\times$10$^{-9}$\\
P$^{+}$      &2.00$\times$10$^{-10}$ \\
   Cl$^{+}$      & 1.00$\times$10$^{-9}$\\
 F    & 2.00$\times$10$^{-9}$\\
ortho-para (o/p) & 10$^{-3}$\\
      \hline
  \normalsize
  \label{tab:initial}
  \end{tabular}
\end{table}

%

To discuss the results obtained from the observations, we have decided to model two types of molecular clouds defined dividing the sample 
of sources in two sub-groups:
\begin{itemize}
\item Cold Model (CM): defined from the four sources with a $T_{\rm dust}$ of about 10 K (see right panel of Fig.~\ref{fig:rapcoldens-varie}). 
These sources are the HMSCs G034--G2, G034--F1, G034--F2, and G028--C1;
\item Warm Model (WM): defined from all of the other sources, which have $T_{\rm dust}$>10~K.
 \end{itemize}

For both chemical model types we fixed the cosmic-ray ionization rate ($\zeta$), the visual extinction ($A_{\rm V}$), the grain albedo ($\omega$), 
the grain radius ($a_{\rm g}$), the grain material density ($\rho_{\rm g}$), the ratio between the diffusion and the binding energy of a species on 
dust grains ($\epsilon$), and the dust-to-gas mass ratio ($R_{\rm g}$) to the values given in Table \ref{tab:table-physpar}. The gas temperature 
($T_{\rm gas}$) and total number density of H nuclei\footnote{$n_{\rm H}$ = $n$(H) + 2$n({\rm H}_{2})\simeq 2n({\rm H}_{2})$ in dense molecular 
clouds like those simulated in this work.} ($n_{\rm H}$) of the two models have been defined from the average $T_{\rm dust}$ and $n$(H$_{2}$) 
of the two sub-groups of sources, respectively. We have decided to use the dust temperatures, instead of the kinetic temperatures, because they 
were all derived with the same method (Mininni et al., submitted), while the kinetic temperatures were not, and assumed 
$T_{\rm gas} = T_{\rm dust}$ for simplicity. For the cold model $T_{\rm gas}$ = 10~K and $n_{\rm H}$ = 3.4$\times$10$^{4}$ cm$^{-3}$, 
and for the warm model $T_{\rm gas}$ = 27 K and $n_{\rm H}$ = 2.4$\times$10$^{5}$ cm$^{-3}$.
We adopt as initial conditions the same used by \citet{colzi20} with an ortho-para (o/p) ratio of 10$^{-3}$.
Since in warm regions ($\> 20$~K) the o/p ratio could be higher, and because N chemistry may be affected by it, 
we tried also to assume a o/p ratio for the WM of $10^{-2}$ and of $10^{-1}$. This test was done with a fixed temperature of 
27 K and we do not take into account the possibility that the o/p ratio could be even higher while the temperature increases. 
In both cases, the results that will be discussed in the next section do not change. 

It has to be noted that for this work, both for the observations and for the chemical modelling we discard any possible effect from 
chemical isotopic fractionation of carbon, which could modify the observed and modelled abundances of HCN and HCO$^{+}$ 
at most by a factor three (see Colzi et al.~\citeyear{colzi20}).
Gas-grain chemistry is followed in our chemical models and it is important in the differentiation between the warm and cold 
sources, as in the former ones CO can be mainly maintained in the gas phase (27 K is indeed just above the sublimation 
temperature of CO), thus contributing to the larger HCO+ abundances in warm regions.

\begin{table}
\caption{Values of the physical parameters fixed in each model.}
  \begin{tabular}{ll}
  \hline
  Parameter   & Value\\ 
  \hline
 $\zeta$  & 1.3$\times$10$^{-17}$ s$^{-1}$\\
  $A_{\rm V}$   & 10 mag \\
  $\omega$& 0.6\\
  $a_{\rm g}$ & 10$^{-5}$ cm\\
  $\rho_{\rm g}$& 3 g cm$^{-3}$ \\
  $\epsilon= E_{\rm diff}/E_{\rm b}$ &0.6 \\
  $R_{\rm d}$= dust-to-gas mass ratio & 0.01 \\
  $T_{\rm gas}$ & 10~K (CM) \\
                        & 27~K (WM) \\
  $n_{\rm H}$  & $3.4\times10^{4}$ cm$^{-3}$ (CM) \\
                       & $2.4 \times10^{5}$ cm$^{-3}$ (WM) \\
                       
  \hline
  \normalsize
  \label{tab:table-physpar}
  \end{tabular}
\centering
\end{table}
 
 \subsection{Model predictions}
\label{predictions}

\begin{figure*}
\centering
\includegraphics[width=45pc]{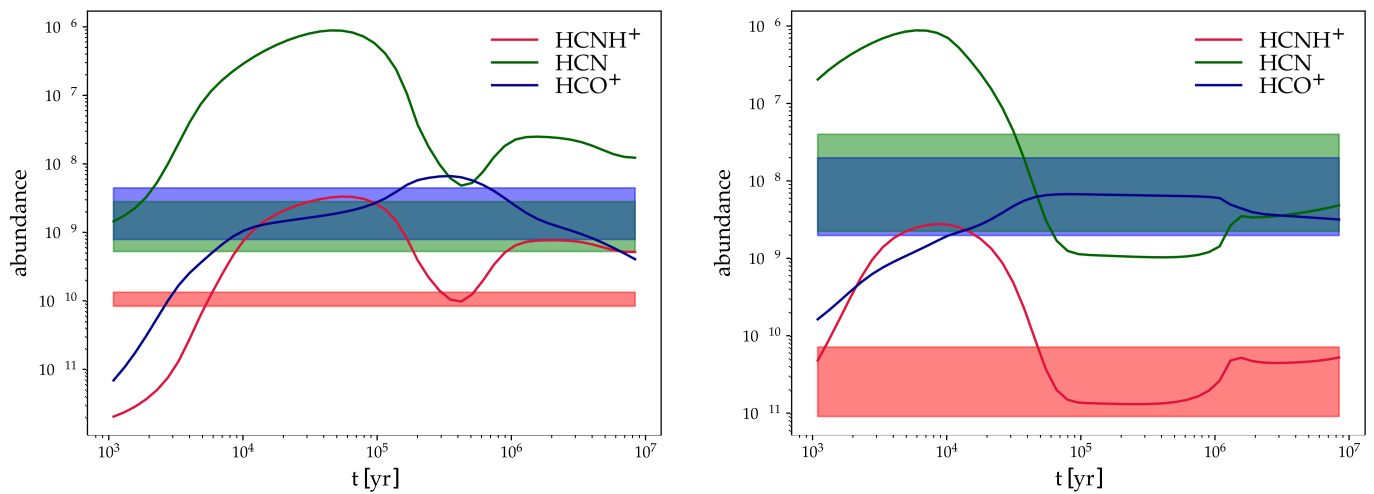}
\caption{Time evolution of HCNH$^{+}$, HCN, and HCO$^{+}$ abundances with respect to H$_{2}$ for the CM (\emph{left panel}) and for the WM (\emph{right panel}). In both panels the coloured horizontal areas represent the range of observed abundances, for HCNH$^{+}$ in red, for HCN in green, and for HCO$^{+}$ in blue, for the two sub-groups of sources.}
\label{fig-abu}
\end{figure*}

\begin{figure}
\centering
\includegraphics[width=23pc]{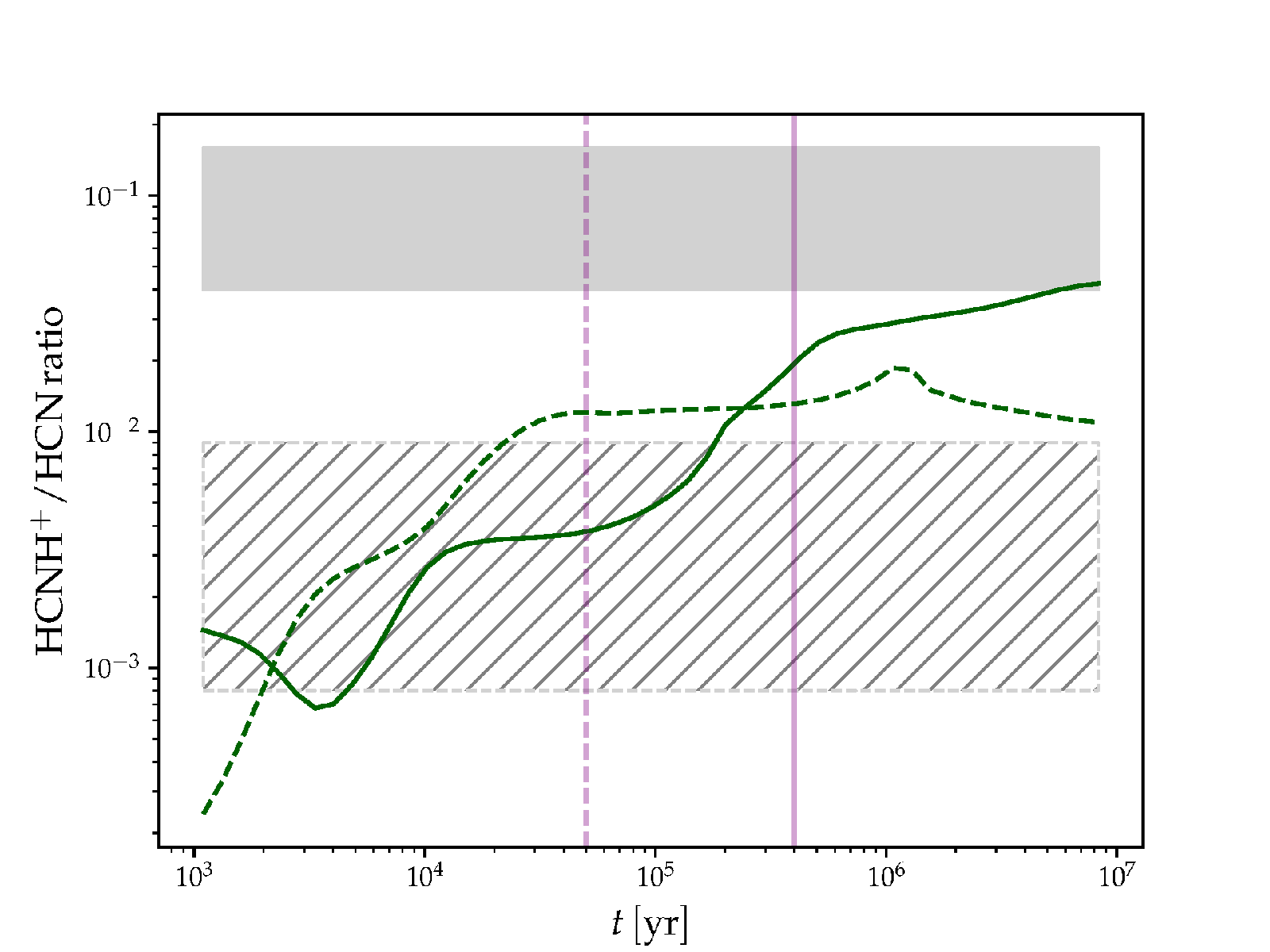}
\caption{HCNH$^{+}$/HCN ratio predicted for the CM (green solid line), and for the WM (dashed green line). The grey coloured areas represent the range of observed HCNH$^{+}$/HCN ratios for the CM (filled area), and for the WM (striped area). The purple vertical lines indicates the times in which the abundances of HCNH$^{+}$, HCN and HCO$^{+}$ are reproduced by the CM ($t_{\rm CM}$=4$\times$10$^{5}$ yr, solid line), and by the WM ($t_{\rm WM}$=5$\times$10$^{4}$ yr, dashed line).}
\label{fig-ratio}
\end{figure}

\begin{figure}
\centering
\includegraphics[width=9.2cm]{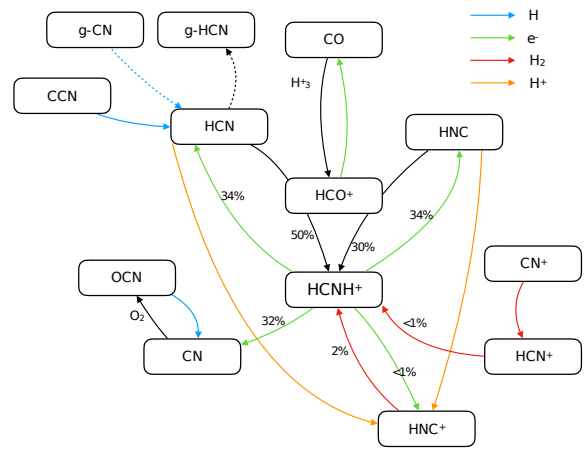}
\caption{Scheme of the chemical network that connects the main molecular species investigated in this work
for the Warm Model (WM in Sect.~\ref{chem-model}). The prefix {\it g-} indicates molecules in icy mantles on dust grains. 
The arrows are colour-coded 
depending on the involved reactant: H (blue), H$_2$ (red), free electrons (green), H$^+$ (red), other species 
(black; the reactant is indicated on the side). Dashed arrows refer to transitions from solid to gas phase, 
or vice-versa. For the reactions forming and destroying \HCNHp, we also indicate their relative contribution 
in percentage.}
\label{fig:elena-plot}
\end{figure}

Figure \ref{fig-abu} shows the predicted abundances for HCNH$^{+}$, HCN, and HCO$^{+}$. These predictions are plotted together 
with the range of the observed abundances, derived as explained in Sect.~\ref{coldens}. For the CM the three abundances can be partially 
reproduced at about $t_{\rm CM}$=4$\times$10$^{5}$ yr. For the WM this happens at two times, 5$\times$10$^{4}$ yr and 3$\times$10$^{6}$ yr, 
but we have decided to take into account for the discussion only the earlier one ($t_{\rm WM}$=5$\times$10$^{4}$ yr), which is more representative 
of the life-time of massive star-forming regions (e.g.~Motte et al.~\citeyear{motte18}). Moreover, at these times the abundance of HCN is 
equal to that of HCO$^{+}$, as it is observed towards the sources studied in this work (see Figs.~\ref{fig:coldens} and \ref{fig:coldens-HCO+}).

At these time scales most of the HCN and HCNH$^{+}$ are already formed through the early-time chemistry (see e.g. Hily-Blant et al.~\citeyear{hilyblant10}), 
HCN being mainly formed from:
\begin{equation}
 {\rm CH}_{2} + {\rm N} \rightarrow {\rm HCN} + {\rm H},
\end{equation}
and HCNH$^{+}$ from:
\begin{equation}
{\rm HCN} + {\rm H}_{3}^{+} \rightarrow {\rm HCNH}^{+} +{\rm H}_{2}.
\end{equation}
Indeed, the abundances of both HCN and \HCNHp\ increase by one 
order of magnitude at chemical times of 10$^{4}$ yr (Fig.~\ref{fig-abu}).

As shown in Fig.~\ref{fig-ratio}, the observed HCNH$^{+}$/HCN ratio can not be reproduced by our chemical models 
at time scales in which the abundances are reproduced. In fact, the predicted HCNH$^{+}$/HCN ratio at $t_{\rm CM}$ is $\sim$0.02, 
lower than the observed ratios, and at $t_{\rm WM}$ is $\sim$0.01, slightly higher than what observed. However, observations and 
model predictions agree with a higher HCNH$^{+}$/HCN ratio in colder sources with respect to the warmer ones.

 \subsection{Formation and destruction reactions of \HCNHp}
 \label{reactions}
 
To perform a detailed analysis and discuss the observed differences between the two sub-groups of sources, i.e. cold/early
starless cores and warmer/evolved objects, we have studied 
in detail the main reactions of formation and destruction which involve HCN, HCNH$^{+}$ and related chemical species. In particular, 
we have taken into account the $t_{\rm CM}$ and $t_{\rm WM}$ times. HCO$^{+}$ is in both models formed mainly 
from CO + H$_{3}^{+}$ and mainly destroyed by HCO$^{+}$ dissociative recombination. However, as will be discussed later, its 
presence is very important in the cycle of reactions involved in the warm model.

Regarding the cold model, HCNH$^{+}$ is mainly formed via:
\begin{equation}
\label{eq-hcnh+-form}
{\rm HCN}^{+} + {\rm H}_{2} \rightarrow {\rm HCNH}^{+} + {\rm H},
\end{equation}
\begin{equation}
\label{eq-hcnh+-form2}
{\rm HNC}^{+} + {\rm H}_{2} \rightarrow {\rm HCNH}^{+} + {\rm H},
\end{equation}
and 
\begin{equation}
{\rm NH}_{3} + {\rm C}^{+} \rightarrow {\rm HCNH}^{+} + {\rm H},
\end{equation}
and mainly destroyed by dissociative recombination forming HCN, HNC and CN with almost the same probability (33.5\%, 33.5\% and 33\%, respectively, Semaniak et al.~\citeyear{semaniak01}):
\begin{equation}
\label{eq-disrec-hcn}
{\rm HCNH}^{+} + e^{-} \rightarrow {\rm HCN} + {\rm H},
\end{equation}
\begin{equation}
{\rm HCNH}^{+} + e^{-} \rightarrow {\rm HNC} + {\rm H},
\end{equation}
and
\begin{equation}
{\rm HCNH}^{+} + e^{-} \rightarrow {\rm CN} + {\rm H} + {\rm H}.
\end{equation}
Moreover, HCN$^{+}$ is mainly formed via CN + H$_{3}^{+}$ closing this part of the cycle of reactions (Woon \& Herbst~\citeyear{weh09}).
Another cycle of reactions involves HCN which at $t_{\rm CM}$ is mainly formed via reaction \eqref{eq-disrec-hcn} and mainly destroyed by 
the presence of HCO$^{+}$, C$^{+}$ and H$_{3}^{+}$. 
All of this is in agreement with the predictions of \citet{hilyblant10}, and \citet{loison14} for low-temperature pre-stellar cores and dark 
molecular clouds, respectively. More recently, for HCNH$^{+}$, this was also confirmed by \citet{quenard17} towards the pre-stellar core L1544.

The main novelty of this work is the study of the main chemical reactions for these chemical species in an environment warmer and denser 
than in the previous studies. In fact, from the WM we have found that HCNH$^{+}$ is mainly formed via:
\begin{equation}
\label{eq-hcn-hco+}
{\rm HCN/HNC} + {\rm HCO}^{+} \rightarrow {\rm HCNH}^{+} + {\rm CO},
\end{equation}
instead of reactions \eqref{eq-hcnh+-form} and \eqref{eq-hcnh+-form2}. The latter reactions are still occurring, but are not efficient any more 
for the formation of protonated HCN because the abundances of HCN$^{+}$ and HNC$^{+}$ decrease by three orders of magnitude with 
respect to the CM. 
The cycle of main reactions for the warm model is shown in Fig.~\ref{fig:elena-plot}.
These chemical pathways are in agreement with the tentative correlations found for the FWHM between 
\HCNHp\ and HCN and \HCOp.

\subsection{Discussion of the assumed parameters}
\label{comparison}

\begin{figure*}
\centering
\includegraphics[width=45pc]{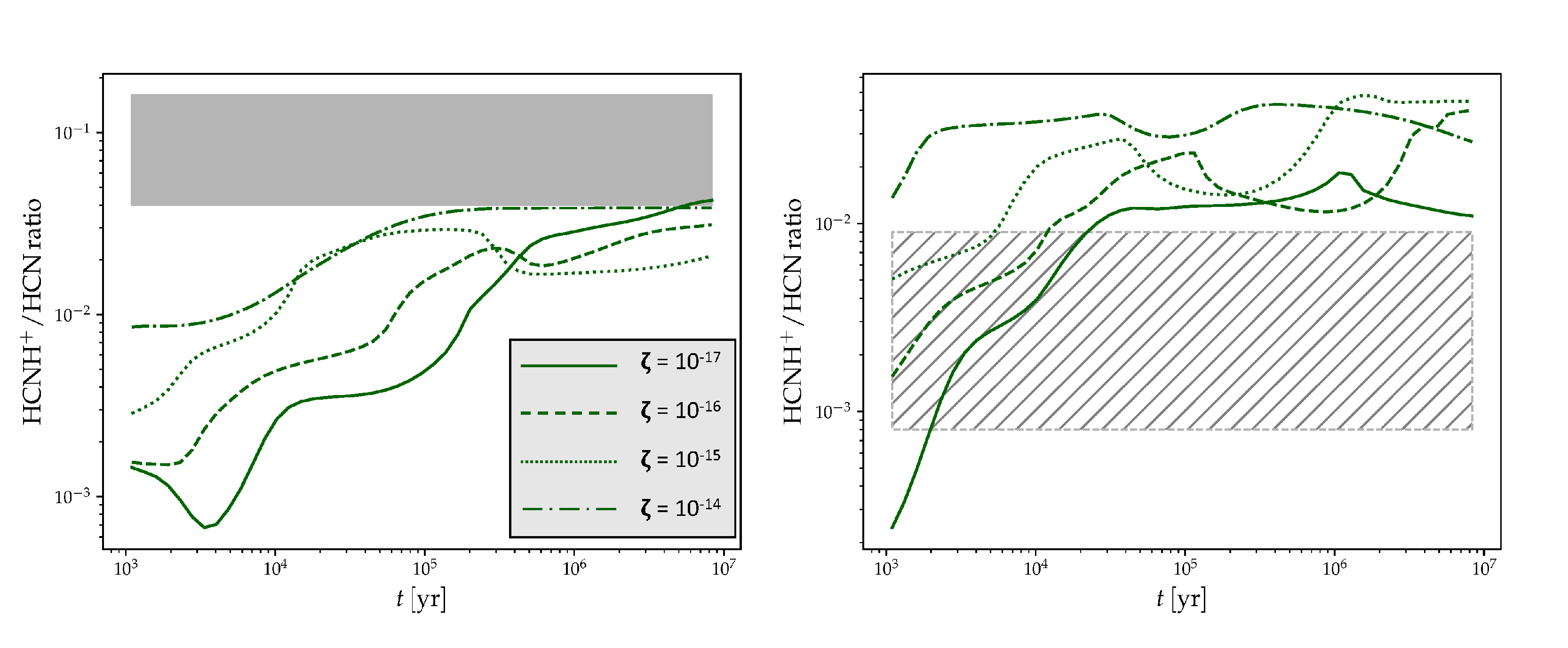}
\caption{HCNH$^{+}$/HCN ratio predicted for the CM (\emph{left panel}), and for the WM (\emph{right panel}), assuming different cosmic-ray ionisation rate, $\zeta$. The grey coloured areas represent the range of observed HCNH$^{+}$/HCN ratios for the CM (filled area in the left panel), and for the WM (striped area in the right panel).}
\label{fig-ratio-zeta}
\end{figure*}

\begin{figure}
\centering
\includegraphics[width=23pc]{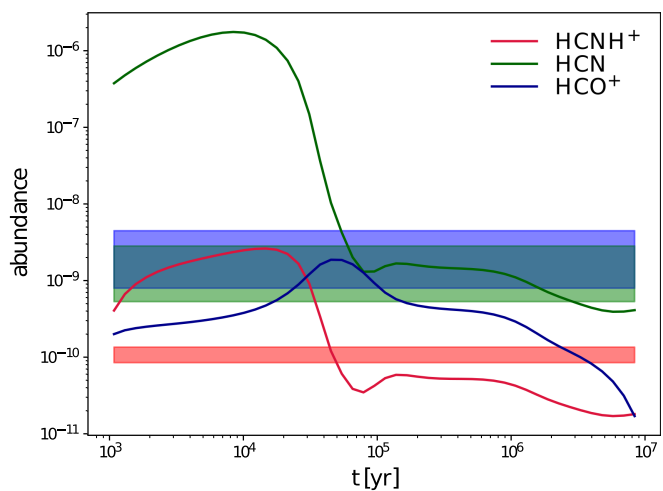}
\caption{Time evolution of HCNH$^{+}$, HCN, and HCO$^{+}$ abundances with respect to H$_{2}$ for the cold model, but assuming a higher $n_{\rm H}$ of 3.4$\times$10$^{5}$ cm$^{-3}$ . The coloured horizontal areas are the same as in the left panel of Fig.~\ref{fig-abu}.}
\label{fig-abundance-cold-10x}
\end{figure}

\begin{figure}
\centering
\includegraphics[width=23pc]{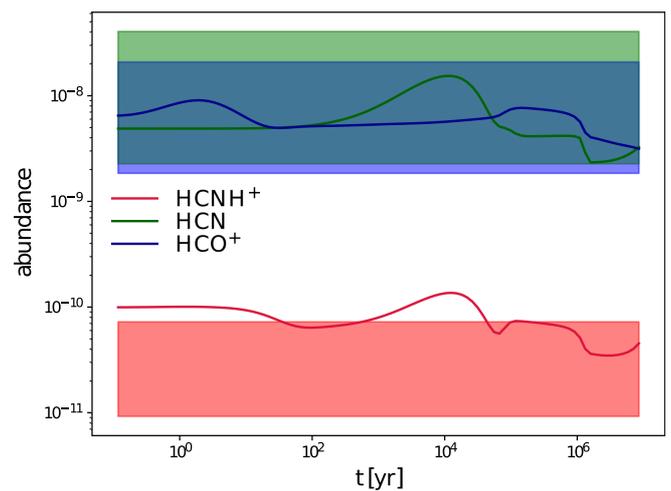}
\caption{Time evolution of HCNH$^{+}$, HCN, and HCO$^{+}$ abundances with respect to H$_{2}$ for the warm model, found assuming as initial conditions the abundances of the CM at $t_{\rm CM}$. The coloured horizontal areas are the same as the right panel of Fig.~\ref{fig-abu}.}
\label{fig-abundance-inicold}
\end{figure}

It is worth noting that the sources have very complex structure, with dense cores embedded in a diffuse envelope, and also with the 
possible presence of extended outflows from the massive protostellar objects. In these models we are simplifying the physical geometry 
of the sources, and most of the parameters are assumed to be constant during the chemical evolution, possibly affecting the predicted 
abundances. Hence, we have checked if changing some of the physical parameters of the two models could lead to a better agreement 
with the absolute HCNH$^{+}$/HCN ratios observed. 

First of all, we tested higher values of the cosmic-ray ionisation rate with respect to the canonical value of 1.3$\times$10$^{-17}$ s$^{-1}$ 
(e.g.~Padovani et al.~\citeyear{padovani09}), of 10$^{-16}$,  10$^{-15}$, and 10$^{-14}$ s$^{-1}$ (e.g. L\'opez-Sepulcre et al.~\citeyear{lopez13}; 
Fontani et al.~\citeyear{fontani17}). We find a higher HCNH$^{+}$/HCN ratio (up to 0.04 already at 10$^{5}$ yr), 
at the lower edge of the observed values, for higher $\zeta$ in the CM (see left panel of Fig.~\ref{fig-ratio-zeta}). In fact, the presence 
of more ions favours the destruction of HCN over the formation of HCNH$^{+}$. Moreover, this trend is the same for the WM, 
making the observed values reproduced for lower $\zeta$ (see right panel of Fig.~\ref{fig-ratio-zeta}). However, these trends are hard to 
understand because the $\zeta$ measured towards HMSCs are lower with respect to those found towards protostellar objects 
(e.g. Fontani et al.~\citeyear{fontani17}). Another discrepancy is that the observed abundances of the molecules could not be reproduced, 
for both models, at times in which the HCNH$^{+}$/HCN ratio matches (see Figs.~\ref{fig:zeta14},~\ref{fig:zeta15}, and~\ref{fig:zeta16}). 
Thus, cosmic-ray ionisation rate could not help us to explain the observed values. 

Secondly, we investigated the possibility of a higher hydrogen nuclei number density for the cold model. In fact, $n_{\rm H}$ has been found 
from the H$_{2}$ column densities, which in turn have been derived in an area with angular dimension 28\asec. However, the emission 
is not resolved and in principle it could mainly come from a smaller region (dense cores inside the high-mass clump). This would lead 
to higher $n_{\rm H}$. Assuming a density of one order of magnitude higher ($n_{\rm H}$ = 3.4$\times$10$^{5}$ cm$^{-3}$) we find 
a better agreement with respect to the observed abundances of HCN, which are lower with respect to the predictions of the initial cold 
model (Fig.~\ref{fig-abundance-cold-10x}). In fact, the higher density, together with the low temperature, makes adsorption onto grain 
surfaces more efficient. However, also the abundance of HCNH$^{+}$ decreases. 

Finally, since high-mass protostellar objects are born in a gas that was previously a starless core, we tried to model the warm sources 
using as initial conditions the abundances of the CM at 4$\times$10$^{5}$ yr, time in which the CM observations are reproduced. 
Interestingly, we found a warm model with almost constant abundances for the three models, well reproduced by the observations 
of HCN and HCO$^{+}$, while slightly above with respect to the observed HCNH$^{+}$ (Fig.~\ref{fig-abundance-inicold}). 
This leads to the same result obtained in the original WM, but spread during the chemical evolution.

To conclude, we have found that a change in the initial conditions or in some of the initial parameters of the chemical model would not 
lead to the absolute observed HCNH$^{+}$/HCN ratios. Many of the relevant reactions in the chemical model are not well constrained, 
and perhaps some important pathways are missing. For example, reactions \eqref{eq-hcnh+-form}, \eqref{eq-hcnh+-form2}, and \eqref{eq-hcn-hco+} 
in the KIDA network are just estimated and laboratory measurements are needed to obtain the correct molecular abundances. 

\section{Summary and conclusions}
\label{conc}

We have presented the first survey of \HCNHp\ $J=3-2$ lines, observed with the IRAM-30 telescope, 
towards 26 high-mass star-forming regions. We report 14 detections and two tentative detections,
for a total detection rate of $\sim 62\%$. The total column densities, \Ntot (\HCNHp), calculated assuming 
optically thin lines and local thermodynamic equilibrium conditions, are in the range $0.5 - 10 \times 10^{14}$ \cmq. 
The abundances of \HCNHp\ with respect to H$_2$ are in the range $0.9 - 14 \times 10^{-11}$, and the highest
values are found towards the coldest HMSCs, for which [\HCNHp] is of the order of $10^{-10}$.
The abundance ratios [\HCNHp]/[HCN] and [\HCNHp]/[\HCOp] are both $\leq 0.01$ in all targets
except towards the four coldest HMSCs. Hence, the dominant formation pathways of \HCNHp\ in 
cold/early and warm/evolved regions are likely different. 
We have run two chemical models, a "cold" one and a "warm" one, to attempt to reproduce our results. 
Besides the different temperature, the two models are also adapted to match as much as possible the 
average physical conditions of the four cold(est) HMSC and the other sources. In particular, the main 
chemical reactions leading to the formation and destruction of \HCNHp\ in the "warm" model are investigated 
in this work for the first time.
Our predictions indicate that indeed HCO$^+$ and HCN/HNC are the dominant progenitor species 
of \HCNHp\ in the warm model, while in the cold one \HCNHp\ is mainly formed by
HCN$^+$ and HNC$^+$. Another important result of this study is that the abundance ratios 
[\HCNHp]/[HCN] and [\HCNHp]/[\HCOp] can be a useful astrochemical tool to discriminate 
between different evolutionary phases in the process of star formation. Naturally,
higher angular resolution observations will allow us to better constrain
precise location and extent of the emitting region of \HCNHp\ in the sources. More transitions
will also help constraining more precisely the excitation conditions, both crucial elements
to define better the range of physical parameters appropriate to model the chemistry.

\begin{acknowledgements}
We thank the anonymous Referee for their valuable and constructive comments.
F.F. is grateful to the IRAM-30m staff for their precious help during the observations.
L.C. acknowledges financial support through Spanish grant ESP2017-86582-C4-1-R (MINECO/AEI). 
L.C. also acknowledges support from the Comunidad de Madrid through the Atracci\'on de Talento 
Investigador Modalidad 1 (Doctores con experiencia) Grant (COOL: Cosmic Origins Of Life; 2019-T1/TIC-15379; PI: Rivilla).
The  research  leading  to  these results has received funding from the European Commission Seventh 
Framework Programme (FP/2007-2013) under grant agreement No 283393 (RadioNet3).
\end{acknowledgements}

%
%

\newpage

\renewcommand{\thefigure}{A-\arabic{figure}}
\setcounter{figure}{0}
\section*{Appendix A: model predictions for different $\zeta$}
\label{appa}
We show in this appendix the predictions of our chemical models for $\zeta$ different
from the canonical value $\zeta = 1.3 \times 10^{-17}$ s$^{-1}$, adopted in Fig.~\ref{fig-abu}.

\begin{figure*}
\centering
{\includegraphics[width=17cm,angle=0]{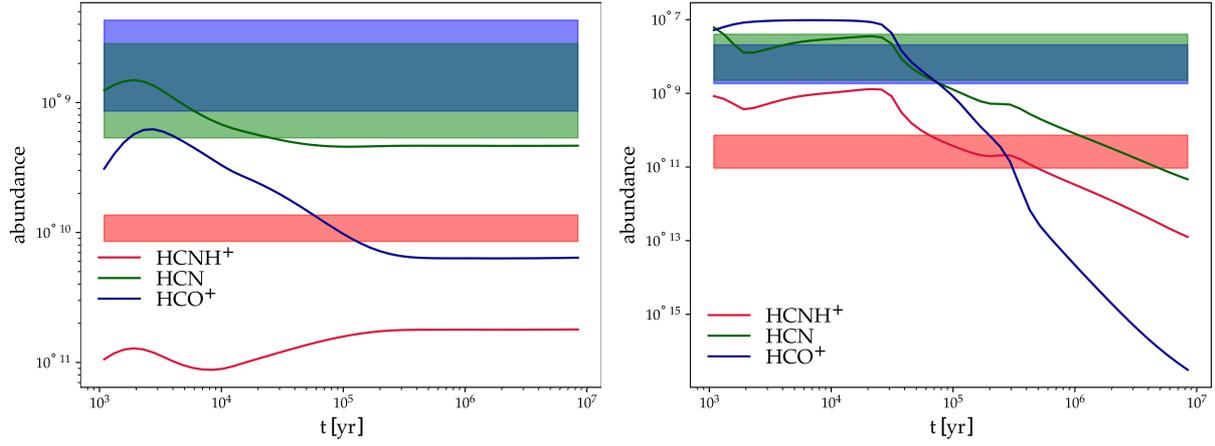}}
\caption{Same as Fig.~\ref{fig-abu} assuming a cosmic-ray ionisation rate $\zeta = 1.3 \times 10^{-14}$ s$^{-1}$.}
\label{fig:zeta14}
\end{figure*}

\begin{figure*}
\centering
{\includegraphics[width=17cm,angle=0]{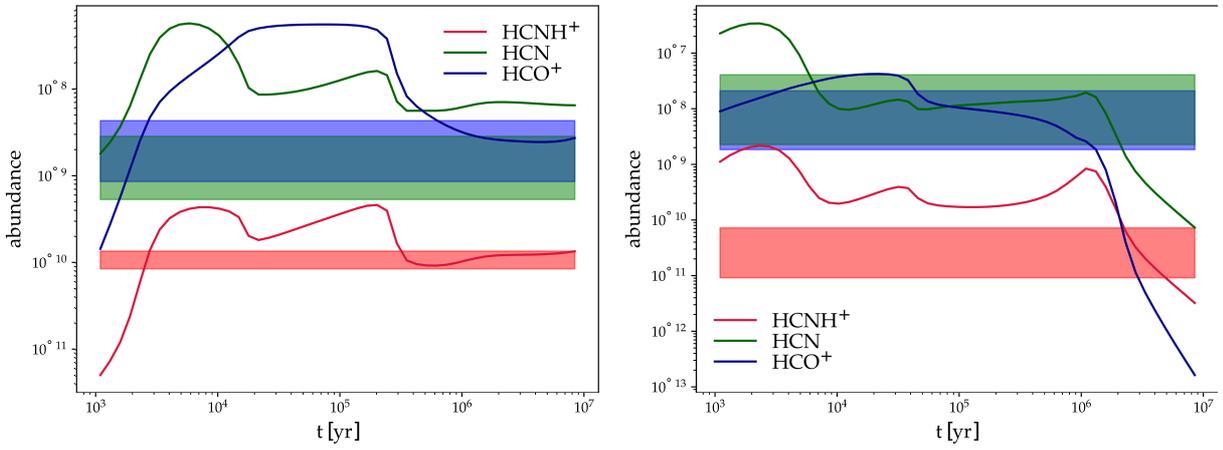}}
\caption{Same as Fig.~\ref{fig:zeta14} for $\zeta = 1.3 \times 10^{-15}$ s$^{-1}$.}
\label{fig:zeta15}
\end{figure*}

\begin{figure*}
\centering
{\includegraphics[width=17cm,angle=0]{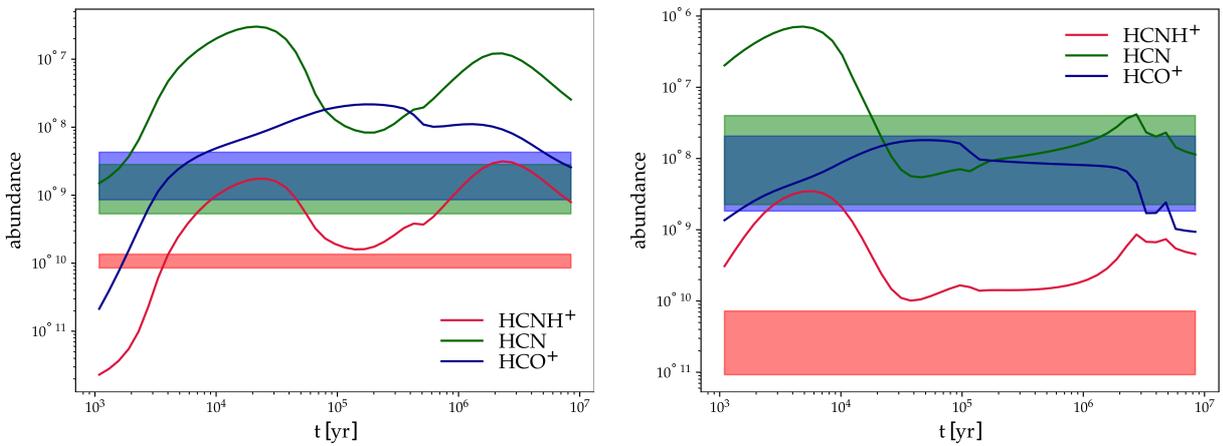}}
\caption{Same as Fig.~\ref{fig:zeta14} for $\zeta = 1.3 \times 10^{-16}$ s$^{-1}$.}
\label{fig:zeta16}
\end{figure*}

\end{document}